\documentclass{article}

\usepackage{fullpage}
\usepackage{authblk}

\usepackage[english=american]{csquotes}
\usepackage{graphicx,amsmath,amssymb,amsfonts,amsthm,mathtools} 

\usepackage{subcaption}

\usepackage[hidelinks]{hyperref} 
\usepackage[dvipsnames]{xcolor}
\usepackage[numbers,sort]{natbib}

\usepackage{enumerate} 
\usepackage{booktabs} 
\newcommand{\bigcell}[2]{\begin{tabular}[t]{@{}#1@{}}#2\end{tabular}}
\usepackage{longtable}
\usepackage{ulem} 
\usepackage[width=1\textwidth,labelfont=bf]{caption}
\usepackage[margin=.75in]{geometry}

\newcommand{\BTnstar}{\mathcal{BT}^\ast_n}

\newcommand{\BTn}{\mathcal{BT}_n}
\newcommand{\Tfb}{T^{\mathit{fb}}_h}

\newcommand{\Tcat}{T^{\mathit{cat}}_n}

\newcommand{\cT}{\mathcal T}
\newcommand{\SymAsym}{\circledast}

\newcommand{\myrefSecTBIndices}{2 }
\newcommand{\myrefSecTreeModels}{3 }
\newcommand{\myrefSecPowerResults}{5 }

\newcommand{\myrefRemarkPopgen}{3.1 }
\newcommand{\myrefOtherNullModels}{6 }
\newcommand{\myrefFigABDvsAldous}{20 }

\definecolor{verylightgray}{rgb}{0.95,0.95,0.95}
\definecolor{othergray}{rgb}{0.55,0.55,0.55}

\usepackage{listings}                           
\title{Tree balance in phylogenetic models}
\author[1]{Sophie J. Kersting\thanks{\url{sophie.kersting@uni-greifswald.de}}}
\author[2]{Kristina Wicke \thanks{\url{kristina.wicke@njit.edu}}}
\author[1]{Mareike Fischer\thanks{\url{mareike.fischer@uni-greifswald.de, email@mareikefischer.de}}}

\affil[1]{Institute of Mathematics and Computer Science, University of Greifswald, Greifswald, Germany}
\affil[2]{Department of Mathematical Sciences \protect \\New Jersey Institute of Technology, Newark, NJ, USA}

\date{}

\begin{document}
\maketitle

\begin{abstract}
\noindent Tree shape statistics, particularly measures of tree (im)balance, play an important role in the analysis of the shape of phylogenetic trees. With applications ranging from testing evolutionary models to studying the impact of fertility inheritance and selection, or tumor development and language evolution, the assessment of measures of tree balance is important. Currently, a multitude of at least 30 (im)balance indices can be found in the literature, alongside numerous other tree shape statistics.

This diversity prompts essential questions: How can we assist researchers in choosing only a small number of indices to mitigate the challenges of multiple testing? Is there a preeminent balance index tailored to specific tasks? 

This research expands previous studies on the examination of index power, encompassing almost all established indices and a broader array of alternative models, such as a variety of trait-based models. Our investigation reveals distinct groups of balance indices better suited for different tree models, suggesting that decisions on balance index selection can be enhanced with prior knowledge. Furthermore, we present the \textsf{R} software package \textsf{poweRbal} which allows the inclusion of new indices and models, thus facilitating future research on the power of tree shape statistics.
\end{abstract}

\section{Introduction}
\paragraph{History of tree balance in phylogenetic models}
A key aspect in the study of evolution is to understand the forces that drive speciation and extinction processes and their effect on the evolution of species and taxa of higher level \cite{Futuyma_evolution_2001}. In the 1960s and 1970s, the idea formed that such evolutionary processes can be modeled using stochastic processes (see, e.g.,~\cite{raup_stochastic_1973, Raup_stochastic_1974, Schopf_genomic_1975, Gould_shape_1977, Raup_stochastic_1977, Schopf_evolving_1979,edwards_reconstruction_1964,edwards_estimation_1970}). Around the same time, researchers started assessing the shapes of (phylogenetic) trees via tree shape statistics (mostly tree balance indices, which measure the degree of imbalance/asymmetry in trees). Tree shape statistics are nowadays used in a variety of contexts, ranging from phylodynamics and the study of viruses (e.g.,~\cite{norstrom_phylotempo_2012,colijn_phylogenetic_2014,pompei2012phylogenetic}) to  phylolinguistics and the study of language diversification (e.g.,~\cite{holman_languages_2010}); see also \cite[Chapter 27]{fischer_tree_2023} for further applications of tree shape statistics.   

Returning to the context of macroevolution and the 1970s, the research community also started to analyze the probabilities of certain shape patterns under basic phylogenetic models \cite{farris_expected_1976, harding_probabilities_1971}, and several biogeographic studies introduced the use of null models in macroevolution~\cite{rosen_vicariant_1978, Simberloff_there_1981, Simberloff_calculating_1987}. At the intersection of these topics, a whole research branch emerged, which is concerned with developing stochastic models that are good and useful explanations of diversification based on the analysis of tree shapes (see, e.g., \cite{savage_shape_1983, slowinski_probabilities_1990, kirkpatrick_searching_1993, mooers_tree_1995, mooers_inferring_1997, Aldous_stochastic_2001, blum_statistical_2005, blum_which_2006, Verboom_species_2020, purvis_evaluating_2002, rosen_vicariant_1978, fusco_new_1995, agapow_power_2002}).

Intuitively, when the moments of a tree shape statistic under a given null model are known, or when the null distribution can be approximated through simulation -- or sometimes through asymptotic distributions, statistical methods to test a given tree against the null hypothesis (i.e., that the tree was obtained under the null model) can be devised. These results can then be used to decide whether a given null model is indeed a realistic model for evolution.

Over the last decades, a vast range of tree shape statistics on the one hand, and phylogenetic or macroevolutionary models on the other hand, have been developed. Due to rapid developments in both fields, studies combining both fields, i.e.,  investigating phylogenetic models through tree balance, have been limited in scope to a handful of tree balance indices and models (see, e.g.,~\cite{heard_patterns_1992, kirkpatrick_searching_1993, heard_patternsrate_1996, agapow_power_2002, blum_statistical_2005, heard_shapes_2007, kersting_genetic_2020, kersting_measuring_2021}). As a result, the idea of using tree shape statistics in the study of phylogenetic models has not reached its full potential yet. Indeed, there has been no exhaustive analysis of the power of different tree shape statistics. In other words, there has been no exhaustive study of the question which tree shape statistics are best at detecting and testing specific evolutionary models.

The recent survey of tree balance indices by Fischer et al.~\cite{fischer_tree_2023} has provided the first pillar for this endeavor: a comprehensive list of all established tree balance indices and further tree shape statistics. These range from old and widely used indices such as the Sackin and Colless indices \cite{sackin_good_1972,colless_review_1982}, to fairly new approaches such as the rooted quartet index \cite{coronado_balance_2019} or the symmetry nodes index \cite{kersting_measuring_2021}. 
As part of this manuscript, we provide the second pillar: an extensive overview and list of basic (phylogenetic) tree models and their parameters (see supplementary material \cite{FischerKerstingWickeSupp}), which is more updated and comprehensive than earlier surveys (such as the excellent review by Mooers et al.~\cite{mooers_models_2007}) on this topic. 

\paragraph{Aims of this manuscript} 
The main purpose of our work, though, is to bridge the above-mentioned two pillars  by providing the means to examine the power of all established tree shape statistics for a wide array of models.  While we focus on their power to detect the Yule model against certain discrete-time models for the purpose of this paper, our general framework can easily be extended to other null and alternative models as well as new tree shape statistics. In particular, we introduce the \textsf{R} software package \textsf{poweRbal}, which can be used to perform analyses analogous to the ones presented here. This package currently includes almost all established tree balance indices as well as numerous macroevolutionary models, but is implemented in such a way that new tree shape statistics and new models can easily be incorporated. Detailed instructions on how to use the package are given in the supplementary material~\cite{FischerKerstingWickeSupp}. In general, we suggest using this software package as a precursor to subsequent analyses and research: The user specifies the research question (e.g., which null and alternative models to investigate), and inputs these choices into our software. Our software then provides the user with the most powerful tree shape statistics for this setting to be used in subsequent statistical analyses of the data. This minimizes problems related to multiple testing and ensures that no better (i.e., more powerful) tree shape statistic is overlooked. 

\paragraph{Related work}
As indicated above, our paper is not the first to analyze tree shape statistics in the context of evolutionary models, but previous studies have been less comprehensive in scope. Nonetheless, we briefly summarize these studies and highlight some alternative approaches that have been used in the literature. First, similar to our study, several authors have studied the power of tree shape statistics to detect deviations from a given phylogenetic null model (e.g.,~\cite{heard_patterns_1992, kirkpatrick_searching_1993, heard_patternsrate_1996, agapow_power_2002, blum_statistical_2005, heard_shapes_2007, kersting_genetic_2020, kersting_measuring_2021}). Using the Yule model as the null model, these studies have employed a handful of tree shape statistics and a handful of alternative models in their investigation. While some tree shape statistics such as the Colless index~\cite{colless_review_1982} have been analyzed in several of these studies, others, such as the rooted quartet index~\cite{coronado_balance_2019}, have not been included at all. However, we remark that very recently and independent of our study, Khurana et al.~\cite{khurana_limits_2023} used a large selection of the tree shape statistics categorized in \cite{fischer_tree_2023} in a study aimed at discriminating between empirical trees and trees generated under a constant-rate birth-death model (see Section \myrefOtherNullModels in the supplementary material for a more thorough discussion of the study by Khurana et al.~\cite{khurana_limits_2023}.) Nevertheless, by including all established tree shape statistics as well as a large variety of alternative models in our analyses, we complement and expand the existing literature. 

In addition to analyzing the power of tree shape statistics in the context of macroevolutionary null models, tree shape statistics have also been used in testing for biogeographic null models (see, e.g., \cite[p.~108]{heard_shapes_2007}). Here, the aim is to decide whether a local subphylogeny is significantly more symmetric or asymmetric than expected if a random phylogeny of the same size is picked from the underlying larger tree.

Finally, there have been some approaches of evaluating tree shape statistics outside the realm of classic statistical hypotheses tests. For instance, Matsen~\cite{matsen_geometric_2006} took a geometric approach to quantifying the power of tree shape statistics to differentiate between similar and different trees, where similarity of trees was measured using the so-called nearest neighbor interchange distance. Intuitively, the idea is that under good tree shape statistics, similar trees should receive similar values, whereas distant trees should receive different values. Hayati et al.~\cite{hayati_new_2019} recently followed up on Matsen's idea, albeit using a slightly different approach (for a more detailed discussion of both approaches, see \cite[pp.~5-10]{hayati_new_2019}).

\section{Preliminaries}
Before we can present our results, we introduce some general definitions and notation, where we mainly follow the terminology of \cite{fischer_tree_2023}. Throughout this manuscript, $X$ denotes a non-empty finite set, which is often referred to as a \emph{taxon set} or simply a \emph{species set}. If not stated otherwise, we may assume $X = \{1, 2, \ldots, n\}$. 

\paragraph{Rooted binary phylogenetic trees and related concepts} 
A \emph{rooted binary phylogenetic $X$-tree $\cT$} is a tuple $\cT = (T, \phi)$, where $T=(V(T), E(T))$ is a rooted tree with root $\rho$ and \emph{leaf} set $V_L(T)$ such that each \emph{inner vertex} $v \in V(T) \setminus V_L(T)$ has out-degree 2, and $\phi$ is a bijection from the leaf set $V_L(T)$ to $X$. Intuitively, $\cT$ is obtained from $T$ by assigning labels or species names to the leaves of $T$. In particular, $T$ is an unlabeled (graph-theoretical) tree, whereas $\mathcal{T}$ is a leaf-labeled (phylogenetic) tree.  Furthermore, if $\vert X \vert = n$, we also have $\vert V_L(T) \vert = n$. 
$T$ is called a \emph{rooted binary tree} and is also often referred to as the \emph{topology} or \emph{tree shape} of $\cT$. Note that sometimes in the literature the term \emph{tree shape} is used to refer to the distribution of edge lengths among internal and leaf edges, without specific reference to the \emph{topology} \cite{williamson_genealogy_2002}. However, we use the two terms \emph{topology} and \emph{tree shape} interchangeably as \emph{tree shape statistics} is an established term for functions that assess the topology of trees. We use $\mathring{V}(T)$ to denote the set of inner (or interior) vertices of $T$.
For every $n \in \mathbb{N}_{\geq 1}$, we denote by $\BTnstar$ the set of (isomorphism classes of) all rooted binary trees with $n$ leaves and by $\BTn$ the set of (isomorphism classes of) all rooted binary phylogenetic $X$-trees with $n$ leaves. 

Given a rooted binary tree $T$ and an arbitrary vertex $v \in V(T)$, we denote by $T_v$ the pending (also called pendant) subtree of $T$ rooted in $v$ and use $n_v$ to denote the number of leaves in $T_v$. Let $v \in \mathring{V}(T)$ be an inner vertex of $T$ with children $v_1$ and $v_2$. We call $v$ a \emph{symmetry node} or \emph{symmetry vertex} if $T_{v_1}$ and $T_{v_2}$ are isomorphic, i.e., have the same shape. The number of symmetry nodes of $T$ is denoted as $s(T)$.

\begin{figure}[htbp]
    \centering
    \includegraphics[width=0.95\textwidth]{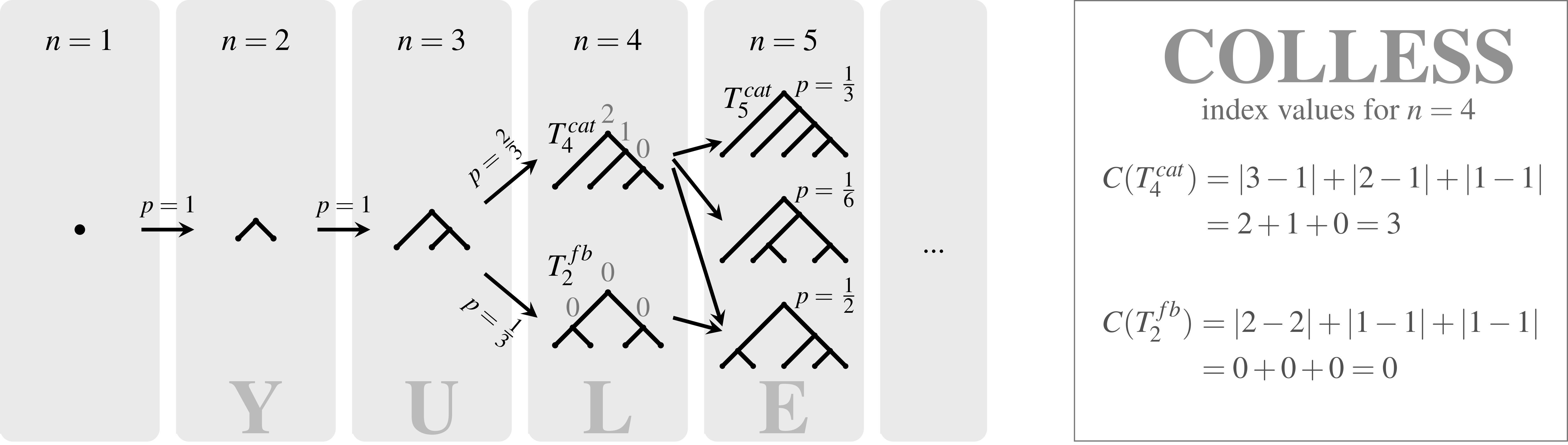}
    \caption{Left: Illustration of the Yule model using the forward process (\enquote{tree-growing}). At each step from $n$ to $n+1$, a leaf is chosen uniformly at random and replaced by a cherry. Thus, under the Yule model, the caterpillar tree on four leaves occurs with probability $P_{Y,4}(T^{cat}_4)=\frac{2}{3}$, whereas the fully balanced tree of height two occurs with probability $P_{Y,4}(T^{fb}_2)=\frac{1}{3}$. Note that only considering $n=4$ leaves might lead to the wrong impression that the Yule model favors highly imbalanced trees. In fact, the caterpillar tree is actually getting less and less likely the more leaves are involved (as only the choice of any of the two leaves of the unique cherry in the caterpillar on $n$ leaves can generate a caterpillar on $n+1$ leaves). In fact, as the figure shows, already for $n=5$, the caterpillar is not the most likely tree anymore. Moreover, note that there are 12 phylogenetic trees of caterpillar shape and 3 phylogenetic trees of fully balanced shape. Thus, under the Yule model, the probability to observe a specific phylogenetic tree $\cT$ of caterpillar shape is $P_{Y,4}(\cT) = \frac{2}{3} \cdot \frac{1}{12} = \frac{1}{18}$, and the probability to observe a specific phylogenetic tree $\cT$ of fully balanced shape is $P_{Y,4}(\cT) = \frac{1}{3} \cdot \frac{1}{3} = \frac{1}{9}$. \\
    Right: Calculation of the Colless index values, i.e., the sum of the absolute differences in the number of leaves in the two maximal pending subtrees of all inner nodes (marked in gray above the respective nodes), for $n=4$. As the Colless index is an imbalance index, more imbalanced trees are assigned higher values.}
    \label{fig:yule_colless}
\end{figure}

Finally, we introduce two special types of trees (see Figure~\ref{fig:yule_colless} for examples). First, the \emph{caterpillar (or comb) tree}, denoted as $\Tcat$, is the unique rooted binary tree which either consists of a singleton leaf or contains precisely one cherry, where a \emph{cherry} is a pair of leaves with a common parent.
The \emph{fully balanced tree}, denoted as $\Tfb$, is the unique rooted binary tree with $n=2^h$ leaves in which all leaves have depth exactly $h$ with $h \in \mathbb{N}_{\geq 0}$. Here, the \emph{depth} $\delta_T(l)$ of a leaf $l$ is the number of edges on the path from the root of $T$ to $l$.

\paragraph{Tree balance indices}
A function $t:\BTnstar\rightarrow\mathbb{R}$ is called a \textit{(rooted) binary tree shape statistic (TSS)}. Note that $t(T)$ depends only on the shape of $T$ and not on the labeling of vertices or the lengths of edges even if TSS in practice are often applied to trees with such properties. A binary tree shape statistic $t$ is called a \emph{balance index} if and only if i) the caterpillar tree $\Tcat$ is the unique tree minimizing $t$ on $\BTnstar$ for all $n\geq 1$ and ii) the fully balanced tree $\Tfb$ is the unique tree maximizing $t$ on $\BTnstar$ for all $n=2^h$ with $h\in\mathbb{N}_{\geq 0}$.
\emph{Imbalance indices} are defined analogously with the extremal trees swapped. 

In Tables~\ref{Table_Balance} and \ref{Table_Imbalance} we list all tree (im)balance indices considered in this manuscript with a short description and their formal definitions.  We refer the reader to the supplementary material~\cite{FischerKerstingWickeSupp} for further details on terminology.

A very intuitive example of how to measure tree imbalance is given by the Colless index I-9, which for each inner node $v$ compares the number of leaves in the two pending subtrees induced by $v$, say $n_{v_1}$ and $n_{v_2}$, and takes the sum over these absolute differences, also known as balance values $bal_T(v)= \vert n_{v_1} - n_{v_2} \vert$ (see Figure \ref{fig:yule_colless} on the right).

\begin{table}[htbp]
\caption{List of tree balance indices (adapted from \cite{fischer_tree_2023}). The first column contains an identifier that helps tracking the individual indices in the figures of Section \ref{sec:results} (an ID in parentheses indicates that the particular index belongs to a group of \enquote{equivalent} indices (see Section \ref{sec:method}) and that another representative, the one with the same ID without parentheses, was used in the analyses). The second column contains the name of the index and a brief description and the third column its formal definition. Details on terminology can be found in the supplementary material~\cite{FischerKerstingWickeSupp}.} 
\label{Table_Balance}
\footnotesize
\begin{tabular}[t]{cll}
\toprule \addlinespace
\multicolumn{2}{l}{\textbf{Balance indices}} & \\ \addlinespace
ID & Index & Definition\\
\midrule \addlinespace
I-1 & \parbox[t]{9.8cm}{\raggedright \textbf{$\boldsymbol{B_1}$ index}~\cite{shao_tree_1990} \\ sum of the reciprocals of the heights of the subtrees of $T$ rooted at inner vertices  of $T$ (except for $\rho$)} & \parbox[t]{6.5cm}{\raggedright $B_1(T) \coloneqq \sum\limits_{v \in \mathring{V}(T) \setminus \{\rho\}} h(T_v)^{-1}$}\\ \addlinespace
I-2 & \parbox[t]{9.8cm}{\raggedright \textbf{$\boldsymbol{B_2}$ index}~\cite{agapow_power_2002,hayati_new_2019,kirkpatrick_searching_1993,shao_tree_1990}\\ Shannon-Wiener information function (measures the equitability of arriving at the leaves of $T$  when starting at the root and assuming equiprobable branching at each inner vertex)} & \parbox[t]{6.5cm}{\raggedright $B_2(T) \coloneqq - \sum\limits_{x \in V_L(T)} p_x \cdot \log(p_x)$}\\ \addlinespace
I-3 & \parbox[t]{9.8cm}{\raggedright \textbf{Furnas rank}~\cite{furnas_generation_1984,kirkpatrick_searching_1993}\\ rank of $T$ according to Furnas' left-light rooted ordering on $\BTnstar$} & \parbox[t]{6.5cm}{\raggedright $F(T) \coloneqq r_n(T)$} \\ \addlinespace
I-4 & \parbox[t]{9.8cm}{\raggedright \textbf{Maximum width}~\cite{colijn_phylogenetic_2014}\\maximum width at any depth} & \parbox[t]{6.5cm}{\raggedright $mW(T) \coloneqq  \max\limits_{i=0,\ldots,h(T)} w_T(i)$} \\ \addlinespace
I-5 & \parbox[t]{9.8cm}{\raggedright \textbf{Maximum width over maximum depth}~\cite{colijn_phylogenetic_2014}\\ratio of maximum width and maximum depth} &\parbox[t]{6.5cm}{\raggedright $W/D(T)$ $ \coloneqq \frac{mW(T)}{mD(T)}$}\\ \addlinespace
I-6 & \parbox[t]{9.8cm}{\raggedright \textbf{Modified maximum difference in widths}~\cite{fischer_tree_2023,colijn_phylogenetic_2014}\\maximum difference in widths of two consecutive depths} & \parbox[t]{6.5cm}{\raggedright $mdelW(T) \coloneqq \max\limits_{i=0,\ldots,h(T)-1} {w_T(i+1)-w_T(i)}$} \\ \addlinespace
I-7 & \parbox[t]{9.8cm}{\raggedright \textbf{Rooted quartet index} for binary trees~\cite{coronado_balance_2019}\\ balance index based on the symmetry of the displayed quartets of $T$} & \parbox[t]{6.5cm}{\raggedright $rQI(T) \coloneqq \vert \{Q\in\mathcal{Q}(T):Q\text{ has shape } T_2^{fb}\} \vert$}\\ \addlinespace
I-8 & \parbox[t]{9.8cm}{\raggedright \textbf{stairs2}~\cite{norstrom_phylotempo_2012,colijn_phylogenetic_2014}\\ratio between the leaf numbers of the smaller and larger pending subtree over all inner vertices} & \parbox[t]{6.5cm}{\raggedright $st2(T) \coloneqq \frac{1}{n-1}\cdot\sum\limits_{v\in\mathring{V}(T)} \frac{\min\{n_{v_1},n_{v_2}\}}{\max\{n_{v_1},n_{v_2}\}}$} \\ \addlinespace
\bottomrule
\end{tabular}
\end{table}

\begin{table}[htbp]
\caption{List of tree imbalance indices (except the Colijn-Plazzotta ranking~\cite{colijn_metric_2018, rosenberg_colijn_2020} for reasons explained in Section \ref{sec:method}; adapted from \cite{fischer_tree_2023}) as well as the cherry index, a simple TSS which does not meet the criteria of an (im)balance index, but is often used as such. The columns are structured similarly to Table \ref{Table_Balance}. Details on terminology can be found in the supplementary material~\cite{FischerKerstingWickeSupp}.}
\label{Table_Imbalance}
\footnotesize
\begin{tabular}[t]{cll}
\toprule \addlinespace
\multicolumn{2}{l}{\textbf{Imbalance indices}} & \\ \addlinespace
ID & Index & Definition\\
\midrule \addlinespace
(I-16) & \parbox[t]{9.5cm}{\raggedright \textbf{Average leaf depth}~\cite{sackin_good_1972, shao_tree_1990}\\mean depth of the leaves of $T$} & \parbox[t]{6.5cm}{\raggedright $\overline{N}(T) \coloneqq \frac{1}{n}\cdot\sum\limits_{x \in V_L(T)} \delta_T(x)$} \\ \addlinespace
(I-16) & \parbox[t]{9.5cm}{\raggedright \textbf{Average vertex depth}~\cite{herrada_scaling_2011}; see also \cite{ford_probabilities_2005,blum_which_2006,hernandez_scaling_2010}\\mean depth of the vertices of $T$} & \parbox[t]{6.5cm}{\raggedright $AVD(T) \coloneqq  \frac{1}{\vert V \vert} \cdot \sum\limits_{v \in V(T)} \delta_T(v)$} \\ \addlinespace
I-9 & \parbox[t]{9.5cm}{\raggedright \textbf{Colless index}~\cite{colless_review_1982} \\ sum of the balance values of the inner vertices of $T$} & \parbox[t]{6.5cm}{$C(T) \coloneqq \sum\limits_{v \in \mathring{V}(T)} bal_T(v) = \sum\limits_{v \in \mathring{V}(T)} \vert n_{v_1} - n_{v_2} \vert$ } \\ \addlinespace
I-10 & \parbox[t]{9.5cm}{\textbf{Colless-like indices}~\cite{mir_sound_2018} \\ sum of the $(D,f)$-balance values of the inner vertices of $T$} & \parbox[t]{6.5cm}{$\mathfrak{C}_{D,f}(T) \coloneqq \sum\limits_{v \in \mathring{V}(T)} bal_{D,f}(v) $} \\ \addlinespace
(I-9) & \parbox[t]{9.5cm}{\raggedright \textbf{Corrected Colless index}~\cite{heard_patterns_1992}\\ $C(T)$ divided by its maximum possible value} & \parbox[t]{6.5cm}{$I_C(T) \coloneqq \frac{2\cdot C(T)}{(n-1)(n-2)}$} \\ \addlinespace
I-11 & \parbox[t]{9.5cm}{\raggedright \textbf{Quadratic Colless index}~\cite{bartoszek_squaring_2021}\\ sum of the squared balance values of the inner vertices of $T$} & \parbox[t]{6.5cm}{ $QC(T) \coloneqq \sum\limits_{v \in \mathring{V}(T)} bal_T(v)^2$}\\ \addlinespace
I-12 & \parbox[t]{9.5cm}{\raggedright \textbf{Equal weights Colless index}~\cite{mooers_inferring_1997}\\ modification of $C(T)$ that weighs the imbalance values of all inner vertices of $T$ equally} & \parbox[t]{6.5cm}{$I_2(T) \coloneqq \frac{1}{n-2}\cdot\sum\limits_{v \in \mathring{V}(T), n_v >2}\frac{bal_T(v)}{n_v-2}$}\\ \addlinespace
I-13 & \parbox[t]{9.5cm}{\raggedright \textbf{Maximum depth}~\cite{colijn_phylogenetic_2014}\\maximum depth of any vertex of $T$; equivalent to $h(T)$} & \parbox[t]{6.5cm}{$mD(T) \coloneqq \max\limits_{v \in V(T)} \delta_T(v) = h(T)$ }\\ \addlinespace
I-14 & \parbox[t]{9.5cm}{\raggedright \textbf{Mean $\boldsymbol{I}$ and $\boldsymbol{I'}$ index}~\cite{fusco_new_1995,purvis_evaluating_2002}\\ mean of the $I_v$ ($I_v'$) values over all inner vertices $v$ with $n_v \geq 4$} & \parbox[t]{6.5cm}{$\overline{I}(T) \coloneqq \frac{1}{\vert \{v \in \mathring{V}(T): n_v \geq 4\} \vert} \cdot  \sum\limits_{v \in \mathring{V}(T), n_v \geq 4} I_v$ \\ $\overline{I'}(T) \coloneqq \frac{1}{\vert \{v \in \mathring{V}(T): n_v \geq 4\} \vert} \cdot  \sum\limits_{v \in \mathring{V}(T), n_v \geq 4} I'_v$}  \\ \addlinespace
I-15 & \parbox[t]{9.5cm}{\raggedright \textbf{Rogers $\boldsymbol{J}$ index}~\cite{rogers_central_1996}\\ number of inner vertices of $T$ which are not perfectly balanced} & \parbox[t]{6.5cm}{$J(T) \coloneqq \sum\limits_{v \in \mathring{V}(T)} (1-\mathcal{I}(bal_T(v)=0))$} \\ \addlinespace
I-16 & \parbox[t]{9.5cm}{\raggedright \textbf{Sackin index}~\cite{sackin_good_1972,shao_tree_1990}\\ sum of the leaf depths of $T$} & $S(T) \coloneqq \sum\limits_{x \in V_L(T)} \delta_T(x)$ \\ \addlinespace
I-17 & \parbox[t]{9.5cm}{\raggedright \textbf{$\boldsymbol{\widehat{s}}$-shape statistic}~\cite{blum_which_2006}\\ sum of $\log(n_v-1)$ over all inner vertices of $T$} & \parbox[t]{6.5cm}{$\widehat{s}(T) \coloneqq \sum\limits_{v \in \mathring{V}(T)} \log(n_v-1)$} \\ \addlinespace
(I-15) & \parbox[t]{9.5cm}{\raggedright \textbf{stairs1}~\cite{norstrom_phylotempo_2012,colijn_phylogenetic_2014}\\proportion of inner vertices that are not perfectly balanced; normalization of Rogers $J$ index} & \parbox[t]{6.5cm}{ $st1(T) \coloneqq \frac{1}{n-1} \cdot \sum\limits_{v \in \mathring{V}(T)} (1-\mathcal{I}(bal_T(v)=0))$} \\ \addlinespace
I-18 & \parbox[t]{9.5cm}{\raggedright \textbf{Symmetry nodes index}~\cite{kersting_measuring_2021}\\ number of inner vertices of $T$ that are not symmetry nodes} & \parbox[t]{6.5cm}{$SNI(T) \coloneqq (n-1)-s(T)$}\\ \addlinespace
I-19 & \parbox[t]{9.5cm}{\raggedright \textbf{Total cophenetic index}~\cite{mir_new_2013}\\ sum of the cophenetic values of all distinct pairs of leaves of $T$} & \parbox[t]{6.5cm}{$\Phi(T) \coloneqq \sum\limits_{\substack{\{x,y\} \in V_L(T)^2 \\ x \neq y}} \varphi_T(x,y)$}\\ \addlinespace
I-20 & \parbox[t]{9.5cm}{\raggedright \textbf{Total $\boldsymbol{I}$ and $\boldsymbol{I'}$ index}~\cite{fusco_new_1995,purvis_evaluating_2002}\\sum/total of the $I_v$ ($I_v'$) values over all inner vertices $v$ with $n_v \geq 4$} & \parbox[t]{6.5cm}{\raggedright $\Sigma I(T) \coloneqq \sum\limits_{v \in \mathring{V}(T), n_v \geq 4} I_v$ \\ $\Sigma I'(T) \coloneqq \sum\limits_{v \in \mathring{V}(T), n_v \geq 4} I'_v$}  \\ \addlinespace
(I-16) & \parbox[t]{9.5cm}{\raggedright \textbf{Total internal path length}~\cite{knuth_volume3_1998}\\ total of the depths of all inner vertices of $T$} & \parbox[t]{6.5cm}{ $TIP(T) \coloneqq \sum\limits_{v \in \mathring{V}(T)} \delta_T(v)$}\\ \addlinespace
(I-16) & \parbox[t]{9.5cm}{\raggedright \textbf{Total path length}~\cite{dobrow_fill_1999,takacs_1992,takacs_1994}\\ total of the depths of all vertices of $T$} & \parbox[t]{6.5cm}{$TPL(T) \coloneqq \sum\limits_{v \in V(T)} \delta_T(v)$} \\ \addlinespace
I-21 & \parbox[t]{9.5cm}{\raggedright \textbf{Variance of leaf depths}~\cite{coronado_sackins_2020, sackin_good_1972,shao_tree_1990}\\variance of the depths of the leaves of $T$} &\parbox[t]{6.5cm}{ $\sigma_N^2(T) \coloneqq \frac{1}n\cdot\sum\limits_{x \in V_L(T)} \left( \delta_T(x) - \overline{N}(T) \right)^2$} \\ \addlinespace
\midrule \midrule \addlinespace
\multicolumn{2}{l}{\textbf{TSS}} & \\ \addlinespace
ID & Index & Definition\\
\midrule \addlinespace
I-22 & \parbox[t]{9.5cm}{\raggedright \textbf{Cherry index}~\cite{mckenzie_distributions_2000}\\ number of cherries in $T$} & \parbox[t]{6.5cm}{\raggedright $ChI(T) \coloneqq c(T)$} \\ \addlinespace
\bottomrule
\end{tabular}
\end{table}

\paragraph{(Phylogenetic) tree models}
A \emph{probabilistic model of binary (phylogenetic) trees $P_n$} (short: (phylogenetic) tree model), with $n \geq 1$, is a family of probability mappings $P_n: \BTnstar \rightarrow [0,1]$ or $P_n: \BTn \rightarrow [0,1]$, respectively, associating a tree $T \in \BTnstar$ or a phylogenetic tree $\mathcal{T} \in \BTn$ with its probability under the model. For some models, explicit formulas for the probability of a particular (phylogenetic) tree are known, while for others this is not the case (yet). Many models implicitly describe the probability distribution by giving an algorithmic process that constructs a (phylogenetic) tree with $n$ leaves under the corresponding model.

Any phylogenetic tree model induces a tree model by setting the probability $P_n(T)$ of a rooted binary tree $T \in \BTnstar$ as the sum of all probabilities $P_n(\mathcal{T})$ over all $\mathcal{T} \in \BTn$ that have $T$ as their underlying topology. Since our aim is to evaluate tree balance indices that work on trees (and do not depend on leaf labels or edge lengths), we focus purely on the (induced) tree models.

A very foundational tree model is the \emph{Yule model}~\cite{yule_mathematical_1925}, which is a pure birth model, where all species have the same constant rate of speciation at all times. To obtain a tree shape, i.e., an unlabeled (rooted binary) tree, under the Yule model using a forward process, starting with a single leaf, a leaf is chosen uniformly at random and replaced by a cherry (see Figure \ref{fig:yule_colless} on the left). Finally, a phylogenetic tree, i.e., a leaf-labeled tree, under the Yule model is obtained by assigning leaf labels uniformly at random to a shape generated by this process. In particular, this means that for the generated (unlabeled) tree shape $T\in \BTnstar$, each of the possible $\frac{n!}{2^{s(T)}}$ (leaf-labeled) phylogenetic trees is equally likely to be chosen. In practice, this can for instance be achieved by choosing, for $i=1\ldots,n$, one leaf at a time uniformly at random from the set of leaves that have not been labeled yet and assigning it label $i$.\footnote{Note that this procedure leads to a multiset of phylogenetic trees in which precisely $2^{s(T)}$ many (isomorphic) versions of each phylogenetic tree are contained (due to the symmetry nodes). Thus, when drawing one phylogenetic tree uniformly at random from this multiset, every phylogenetic tree has the same probability.}
The Yule model plays a central role in phylogenetics and has been used for many decades. It is known under a multitude of names (Yule model, equal-rates-Markov model (ERM), random branching model, Markovian dichotomous branching model, Yule-Harding model, Yule-Harding-Kingman model (YHK), or simply Markovian model) of which we will only use \emph{Yule model} in the remainder of this manuscript. While it can be described under various approaches, the probabilities of obtaining a specific (phylogenetic) tree can be explicitly stated as follows (see, e.g., \cite[Proposition~1]{Steel2001properties} or \cite[Proposition~3.2]{steel_phylogeny_2016}):
\begin{equation} \label{eq:yule_prob}
   P_{Y,n}(\mathcal{T}) = \frac{2^{n-1}}{n!} \cdot \prod\limits_{v \in \mathring{V}(T)} \frac{1}{n_v-1} \qquad \text{and}\qquad P_{Y,n}(T) = 2^{n-1-s(T)} \cdot \prod\limits_{v \in \mathring{V}(T)} \frac{1}{n_v-1}. 
\end{equation}
In particular, the formula for $P_{Y,n}(\cT)$ has been extensively studied in the literature and connections to other fields have been established (see, e.g., \cite{Fuchs2024,Dickey2024,king_mathematical_2023}).

While the Yule model is one of the oldest and most basic tree models, it is often used as a null model in phylogenetic analyses. However, there are many more models available to describe different evolutionary settings. In Tables~\ref{tab:mod_clus_part} to \ref{tab:mod_distrib} we briefly review a collection of tree models. We refer the reader to the supplementary material~\cite{FischerKerstingWickeSupp} for further background and additional models.

\begin{table}[!ht]
    \caption{List of bipartitioning tree models, which iteratively bipartition sets of $m$ leaves into two subsets of $i$ and $m-i$ leaves, respectively. They are defined by a symmetric probability distribution $q_m=(q_m(i); i=1,...,m-1)$ of splitting a given set of size $m$ into two subsets of sizes $i$ and $m-i$.}
    \label{tab:mod_clus_part}
    \setlength\tabcolsep{1.5pt}
    \footnotesize
    \begin{tabular}{llcl}
    \toprule
        \parbox[t]{0.7cm}{ID} & \bigcell{l}{Model} & \bigcell{l}{Probability $q_m(i)$ \\ for $i\in\{1, 2, \ldots, m-1\}$} & Description and examples\\
    \midrule \addlinespace
        M-0 & \parbox[t]{3.5cm}{\raggedright \textbf{YULE} model \cite{yule_mathematical_1925, harding_probabilities_1971}} & \bigcell{l}{$\frac{1}{m-1}$} & \parbox[t]{9.3cm}{\raggedright Given $m$ odd, then splitting it into sizes $1|(m-1)$, $2|(m-2)$,..., or $\lfloor \frac{m}{2}\rfloor|\lceil \frac{m}{2}\rceil$ is equally probable. If $m$ is even, this also holds true, except for the case where $i=m-1=\frac{m}{2}$, in which case the probability is half of that of the other possibilities. See \cite[Equations~(7)~\&~(8)]{slowinski_probabilities_1990}.} \\ \addlinespace
        M-1 & \parbox[t]{3.5cm}{\raggedright \textbf{ALDOUS' $\beta$}-splitting model,  $\beta \geq-2$ \cite{aldous_random_1996}} & \bigcell{l}{$\frac{1}{a_m(\beta)} \frac{\Gamma(\beta + i +1) \Gamma(\beta + m - i +1)}{\Gamma(i+1) \Gamma(m-i+1)}$ \\ with gamma function \\$\Gamma(t) = \int\limits_0^\infty x^{t-1} e^{-x} dt$} & \parbox[t]{9.3cm}{\raggedright Balance increases with $\beta$. Here, $a_m(\beta)$ is a normalizing constant to ensure $\sum_{i=1}^{m-1} q_m(i)=1$. Equals Yule for $\beta=0$, PDA for $\beta=-3/2$, and comb for $\beta=-2$.} \\
        \addlinespace
    \bottomrule
    \end{tabular} 
\end{table}

\begin{table}[!ht]
    \caption{List of \emph{discrete-time tree-growing models without extinction}. These models act at the current leaves and modify their speciation rates. The process starts with a single leaf (the root) with speciation rate $\lambda_0$ and at each time step one of the currently existing leaves, say $p$, is chosen for a speciation event, and split into two leaves, $c_1$ and $c_2$. We denote the resulting rates of $c_1$ and $c_2$ as $(\lambda_{c_1},\lambda_{c_2})$. $\lambda_{p}$ denotes the rate of $p$, whereas $\lambda_i$ denotes the rate of all other leaves not involved in the speciation event. We indicate the corresponding rates for the children and all other vertices in the columns labeled \enquote{Children} and \enquote{Other}.
    Several models are \emph{trait-inspired} or \emph{-based}, i.e., a leaf's speciation rate (in-)directly depends on some quantitative trait $x$. The trait can be age, time, a (partly) inheritable trait (e.g., body size) affecting speciation, environment, number of co-existing species etc. $x_{0}>0$ is the initial trait value of the starting node, $x_{p}$ the parent's value. Again, we indicate the trait values for the children of $p$ and all other nodes in the columns labeled \enquote{Children}, respectively \enquote{Other}. Models that allow the usage of both symmetric and asymmetric speciation (explained in Section~\ref{sec:results}) are marked with $\SymAsym$.
    }
    \label{tab:mod_growing_TS}
    \setlength\tabcolsep{1.75pt}
    \footnotesize
    \begin{tabular}{llccccl}
    \toprule
        \parbox[t]{0.7cm}{ID} & \bigcell{l}{Model} & \bigcell{c}{Children \\$(\lambda_{c_1},\lambda_{c_2})$} & \parbox[t]{1.2cm}{\centering Other \\$\lambda_i$}  & \multicolumn{3}{l}{Description and examples}\\
    \midrule
        M-0 & \parbox[t]{3.5cm}{\raggedright \textbf{YULE} model~\cite{yule_mathematical_1925, harding_probabilities_1971}} & $(\lambda_0,\lambda_0)$ & $\lambda_0$ & \multicolumn{3}{l}{\parbox[t]{9.2cm}{\raggedright Pure birth process, constant and equal rates.}} \\
        \addlinespace
        M-2 & \parbox[t]{3.5cm}{\raggedright Direct-children-only (\textbf{DCO}), $\zeta>0$~\cite[p.~6]{kersting_genetic_2020} $\SymAsym$} & $(\zeta\lambda_0,\zeta\lambda_0)$ & $\lambda_0$ & \multicolumn{3}{l}{\parbox[t]{9.2cm}{\raggedright Factor $\zeta$ has only short-time effect on children. Equals Yule for $\zeta=1$.}} \\
        \addlinespace
        M-3 & \parbox[t]{3.5cm}{\raggedright Inherited fertility (\textbf{IF}), $\zeta>0$~\cite[p.~6]{kersting_genetic_2020} $\SymAsym$} & $(\zeta\lambda_{p},\zeta\lambda_{p})$ & $\lambda_{i}$ & \multicolumn{3}{l}{\parbox[t]{9.2cm}{\raggedright Factor $\zeta$ has long-time and exponential effect. Equals Yule for $\zeta=1$.}} \\
        \addlinespace
        M-4 & \parbox[t]{3.5cm}{\raggedright Unequal fert. inheritance (\textbf{IF-DIFF}), simple biased speciation, $\zeta\geq1$~\cite[p.~1177]{kirkpatrick_searching_1993}} & $\left(\frac{2\zeta}{\zeta+1}\lambda_{p},\frac{2}{\zeta+1}\lambda_{p}\right)$ & $\lambda_{i}$ & \multicolumn{3}{l}{\parbox[t]{9.2cm}{\raggedright Factor $\zeta$ has long-time and exponential effect (opposite effect on the two children). Proportion $\lambda_{c_2}/\lambda_{c_1}=1/\zeta$ with $\lambda_{c_1}+\lambda_{c_2}=2\lambda_{p}$. Equals Yule for $\zeta=1$.}} \\
        \addlinespace
        M-5 & \parbox[t]{3.5cm}{\raggedright \textbf{BIASED SPECIATION}, $\zeta \in [0,1]$, \cite[p.~149]{blum_statistical_2005}} & $(\zeta\lambda_{p},(1-\zeta)\lambda_{p})$ & $\lambda_{i}$ & \multicolumn{3}{l}{\parbox[t]{9.2cm}{\raggedright The parent's rate is divided between its two children according to the given ratio $\zeta$. Note that $\zeta \in [0,0.5]$ yields the same model.}} \\
        \addlinespace
        \midrule
        \multicolumn{7}{l}{\textit{Trait-inspired models}} \\
        \addlinespace
        M-6 & \parbox[t]{3.5cm}{\raggedright Age-step-based fertility (\textbf{ASB}), $\zeta>0$ \cite[p.~6]{kersting_genetic_2020} $\SymAsym$} & $(\lambda_0,\lambda_0)$ & $\zeta\lambda_i$ & \multicolumn{3}{l}{\parbox[t]{9.2cm}{\raggedright Exponential increase ($\zeta>1$) or decrease of rates  ($\zeta<1$) with age (in time steps) of lineage.  Equals Yule for $\zeta=1$.}} \\
        \addlinespace
        M-7 & \parbox[t]{3.5cm}{\raggedright \textbf{SIMPLE BROWNIAN}, $\sigma\geq0$, \cite[p.~6]{kersting_genetic_2020} $\SymAsym$} & \bigcell{c}{$(\lambda_{p} +\varepsilon_{1},$ \\ \quad $\lambda_{p} +\varepsilon_{2})$} & $\lambda_i$ & \multicolumn{3}{l}{\parbox[t]{9.2cm}{\raggedright With $\varepsilon_{1,2}\sim N(0,\sigma^2)$ normally distributed with standard deviation $\sigma$. Rates $<0$ are set to a small value $>0$.}} \\
        \addlinespace
        \midrule
        \multicolumn{2}{l}{\textit{Trait-based models}} & \bigcell{c}{Children \\$(\lambda_{c_1},\lambda_{c_2})$} & \bigcell{c}{Other \\$\lambda_i$}  & \bigcell{c}{Children \\$(x_{c_1},x_{c_2})$} & \bigcell{c}{Other \\$x_i$} & Description\\
        \midrule
        \addlinespace
        M-8 & \parbox[t]{3.5cm}{\raggedright Punctuated (-intermittent) \textbf{LINEAR-BROWNIAN}, (Speciational) Brownian evolution model, trait-based, $\sigma_x,\sigma_\lambda\geq 0$,  \cite[p.~2142]{heard_patternsrate_1996} $\SymAsym$} & \bigcell{c}{$\left(10^{\log_{10}(x_{c_1})+\varepsilon_3}\right.$, \\ $\quad \left.10^{\log_{10}(x_{c_2})+\varepsilon_4}\right)$} & $\lambda_i$ & \bigcell{c}{$(x_{p} +\varepsilon_{1},$ \\ \quad $x_{p} +\varepsilon_{2})$} & $x_i$ & \parbox[t]{5.7cm}{\raggedright Inherited trait influences speciation rate. With $\varepsilon_{1,2}\sim N(0,\sigma^2_x)$ and $\varepsilon_{3,4}\sim N(0,\sigma^2_\lambda)$. Trait values $<0$ are set to a small value $>0$. There is also a \textbf{BOUNDED} version of this model with an upper limit for the trait values, e.g., $x_i\in(0,20)$ symmetrical around $x_0=10$ ($x_0=100$ in \cite{agapow_power_2002}). Equals Yule for $\sigma_\lambda=\sigma_x=0$. } \\
        \addlinespace
        M-9 & \parbox[t]{3.5cm}{\raggedright Punctuated (-intermittent) \textbf{LOG-BROWNIAN}, $\sigma_x,\sigma_\lambda \geq 0$, \cite[p.~2142]{heard_patternsrate_1996} $\SymAsym$   \\ } & \bigcell{c}{$\left(10^{\log_{10}(x_{c_1})+\varepsilon_3}\right.$, \\ $\quad \left.10^{\log_{10}(x_{c_2})+\varepsilon_4}\right)$} & $\lambda_i$ & \bigcell{c}{$\left(10^{\log_{10}(x_p) +\varepsilon_{1},}\right.$ \\ \quad $\left.10^{\log_{10}(x_p) +\varepsilon_{2}}\right)$} & $x_i$ & \parbox[t]{5.7cm}{\raggedright Inherited trait (with proportional changes) influences speciation rate. With $\varepsilon_{1,2}\sim N(0,\sigma^2_x)$ and $\varepsilon_{3,4}\sim N(0,\sigma^2_\lambda)$. Equals Yule for $\sigma_\lambda=\sigma_x=0$. } \\
        \addlinespace
    \bottomrule
    \end{tabular} 
\end{table}

\begin{table}[!ht]
    \caption{List of \textit{discrete-time tree-growing models with extinction}. These models act at the current leaves and modify their speciation and extinction rates. The process starts with a single leaf (the root) with speciation rate $\lambda_0$ and extinction rate $\mu_0$. At each time step one of the currently existing leaves is chosen for a speciation or extinction event. In case of speciation, it is split into two leaves, $c_1$ and $c_2$. We denote the resulting rates of $c_1$ and $c_2$ as $(\lambda_{c_1},\lambda_{c_2})$ and $(\mu_{c_1},\mu_{c_2})$, respectively, and list them in the column labeled \enquote{Children}. 
    $\lambda_i$ and $\mu_i$ denote the rates of all other species and are given in the column labeled \enquote{Other}. $\lambda_{e}$ and $\mu_e$ denote the rates of the leaf chosen for an extinction event (Column \enquote{Extinct}).}
    \label{tab:mod_growing2_TS}
    \setlength\tabcolsep{1.75pt}
    \footnotesize
    \begin{tabular}{llcccl}
    \toprule
        \parbox[t]{0.7cm}{ID} & \bigcell{l}{Model} & \bigcell{c}{Children \\$(\lambda_{c_1},\lambda_{c_2})$\\ $(\mu_{c_1},\mu_{c_2})$} & \bigcell{c}{Other \\$\lambda_i$ \\ $\mu_i$} & \bigcell{c}{Extinct \\$\lambda_e$ \\ $\mu_e$} & Description and examples\\
    \midrule
        M-10 & \parbox[t]{4.3cm}{\raggedright (Simple/constant-rate) \textbf{BIRTH-DEATH} model, $\lambda_0>0$, $\mu_0\geq0$ \\\cite{yule_mathematical_1925, raup_stochastic_1973}} & \bigcell{l}{$(\lambda_0,\lambda_0)$ \\ $(\mu_0,\mu_0)$} & \bigcell{l}{$\lambda_0$ \\ $\mu_0$} & \bigcell{l}{- \\ -}  & \parbox[t]{9cm}{\raggedright \enquote{Reconstructed process}: Birth-death process, where extinct taxa/leaves are removed (and degree-2 vertices are suppressed). Constant and equal speciation and extinction rates. Equals Yule for $\mu_0=0$. Distribution on $\BTnstar$ is the same as for Yule even for $\mu_0>0$ \cite[p.~5]{mooers_models_2007}. } \\
        \addlinespace
        M-11 & \parbox[t]{4.3cm}{\raggedright \textbf{ALTERNATIVE BIRTH-DEATH} model, $\lambda_0>0$, $\mu_0\geq0$} & \bigcell{l}{$(\lambda_0,\lambda_0)$ \\ $(\mu_0,\mu_0)$} & \bigcell{l}{$\lambda_i$ \\ $\mu_i$} & \bigcell{l}{$0$ \\ $0$}  & \parbox[t]{9cm}{\raggedright Extinct taxa/leaves remain in the tree as \enquote{fossils} with speciation and extinction rate 0. Note that, as opposed to other models employing fossils in which fossils are ignored and only extant species are counted, the fossils in this model count towards the total number $n$ of leaves. Equals Yule for $\mu_0=0$.} \\
        \addlinespace
        \midrule
        \multicolumn{6}{l}{\textit{Trait-based models}} \\
        M-12 & \parbox[t]{4.3cm}{\raggedright \textbf{DENSITY}-dependent model \\ \cite{harvey_phylogenies-without-fossils_1994}} & \bigcell{l}{$(\lambda_0,\lambda_0)$ \\ $(\mu_N,\mu_N)$} & \bigcell{l}{$\lambda_0$ \\ $\mu_N$} & \bigcell{l}{- \\ -} & \parbox[t]{9cm}{\raggedright $\mu_N = \begin{cases} \frac{\lambda_0 N}{N^\ast} &\text{if } N < N^\ast\\ \lambda_0 &\text{if $N \geq N^\ast$.} \end{cases}$, where $N$ denotes the number of currently existing lineages, and $N^\ast$ is an equilibrium number of lineages. In other words, the speciation rate is constant, but the extinction rate increases linearly with the number of co-existing lineages until the equilibrium $N^\ast$ is reached at which point $\mu = \lambda_0$. 
        Extinct taxa are removed (reconstructed process).} \\ 
        \addlinespace
        M-13 & \parbox[t]{4.3cm}{\raggedright \textbf{SSE} models \\ state-dependent speciation \\ and extinction models \\ (see, e.g., \cite{maddison_BiSSE_2007, fitzjohn_QuaSSE_2010, fitzjohn_MuSSE_2012, goldberg_GeoSSE_2011,freyman_ChromoSSE_2017, vasconcelos_HiSSE_2022})} & \multicolumn{4}{l}{\parbox[t]{12.5cm}{\raggedright Family of models extending birth-death models to account for speciation, extinction, and trait evolution. Each species is associated with a specific trait/state (observed or hidden), and can transition to another state, or speciate (resp. go extinct) with rates depending on its current state. }}\\
        \addlinespace
    \bottomrule
    \end{tabular} 
\end{table}

\begin{table}[!ht]
    \caption{List of tree-growing models that do not use rates.}
    \label{tab:mod_growEV_ts}
    \setlength\tabcolsep{2pt}
    \footnotesize
    \begin{tabular}{lll}
    \toprule
        \parbox[t]{0.75cm}{ID} & Model  & Description\\
    \midrule
        M-14 & \parbox[t]{4.3cm}{\raggedright \textbf{FORD'S $\alpha$}-model, $\alpha \in [0,1]$ \cite{ford_probabilities_2005, kaur_distributions_2023}}   & \parbox[t]{12.5cm}{Start with a cherry and add an additional temporary root edge $(\rho_{temp},\rho)$. Then, in each step -- let $m$ be the current leaf number -- choose any edge, where each of the $m-1$ inner edges has probability $p=\frac{\alpha}{m-\alpha}$ and each of the $m$ pendant edges has probability $p=\frac{1-\alpha}{m-\alpha}$, subdivide it, and add a new pendant edge with a leaf in each step. Do this until reaching $n$ leaves. Last, delete the edge incident to $\rho_{temp}$. Equals Yule for $\alpha=0$, PDA for $\alpha=1/2$, comb for $\alpha=1$.} \\
        \addlinespace
        M-15 & \parbox[t]{4.3cm}{\raggedright Uniform model, proportional-to-distinguishable-arrangements (or -types) model (\textbf{PDA}) \cite{rosen_vicariant_1978}}  & \parbox[t]{12.5cm}{Start with an unrooted tree with 3 leaves. Then, in each step, choose any edge, subdivide it, and add a new pendant edge with a leaf until reaching  $n$ leaves. Last, choose an edge to insert a root \cite[p.~8]{paradis_enumeration_2023}. }\\
        \addlinespace
    \bottomrule
    \end{tabular} 
\end{table}

\begin{table}[!ht]
    \caption{List of distribution-based tree models. These models generate trees according to some probability distribution. The PDA model here is the same as in Table \ref{tab:mod_growEV_ts}.}
    \label{tab:mod_distrib}
    \setlength\tabcolsep{3.5pt}
    \footnotesize
    \begin{tabular}{llcl}
    \toprule
        \parbox[t]{0.7cm}{ID} & Model & Probabilities & Description\\
    \midrule
        M-15 & \parbox[t]{4.3cm}{\raggedright Uniform model, proportional-to-distinguishable-arrangements (or -types) model (\textbf{PDA}) \cite{rosen_vicariant_1978}} & \bigcell{l}{$P_n(T) = P_n(\mathcal{T}) \cdot \left\vert \mathcal{T}_T\right\vert$ \\ \phantom{$P_n(T)$} $= \frac{n!}{2^{s(T)}\cdot (2n-3)!!}$} & \parbox[t]{8.5cm}{\raggedright All phylogenetic trees $\mathcal{T}\in \BTn$ are equiprobable, i.e., $P_n(\mathcal{T}) = \frac{1}{\vert\BTn\vert}= \frac{1}{(2n-3)!!}$.} \\
        \addlinespace
        M-16 & \parbox[t]{4.3cm}{\raggedright Uniform model, equiprobable-types-model (\textbf{ETM})} & \bigcell{l}{$P_n(T) = \frac{1}{\vert\BTnstar\vert}=\frac{1}{we(n)}$} & \parbox[t]{8.5cm}{\raggedright All rooted binary trees $T\in \BTnstar$ are equiprobable.} \\
        \addlinespace
    \bottomrule
    \end{tabular} 
\end{table}

\clearpage
\section{Methods: How to measure the power of balance indices} \label{sec:method}
While there are various ways of comparing tree balance indices, in the present manuscript, we focus on comparing their power to distinguish alternative phylogenetic models from a null model $P^0_n$. As most previous studies (e.g.,~\cite{heard_patterns_1992, kirkpatrick_searching_1993, heard_patternsrate_1996, agapow_power_2002, blum_statistical_2005, heard_shapes_2007, kersting_genetic_2020, kersting_measuring_2021}) we choose the Yule model M-0 as the null model for the main paper (we present some additional results using other null models in the supplementary material~\cite{FischerKerstingWickeSupp}). The Yule model also belongs to several parameterized tree model families, which allows us to observe the power of TSS as we gradually move away from the Yule model by changing the respective parameter(s). The null hypothesis $H_0$ is that the trees have been constructed under the null model. 

\begin{figure}[ht]
    \centering
    \begin{subfigure}[t]{0.45\textwidth}
    \centering
    \includegraphics[width=\textwidth]{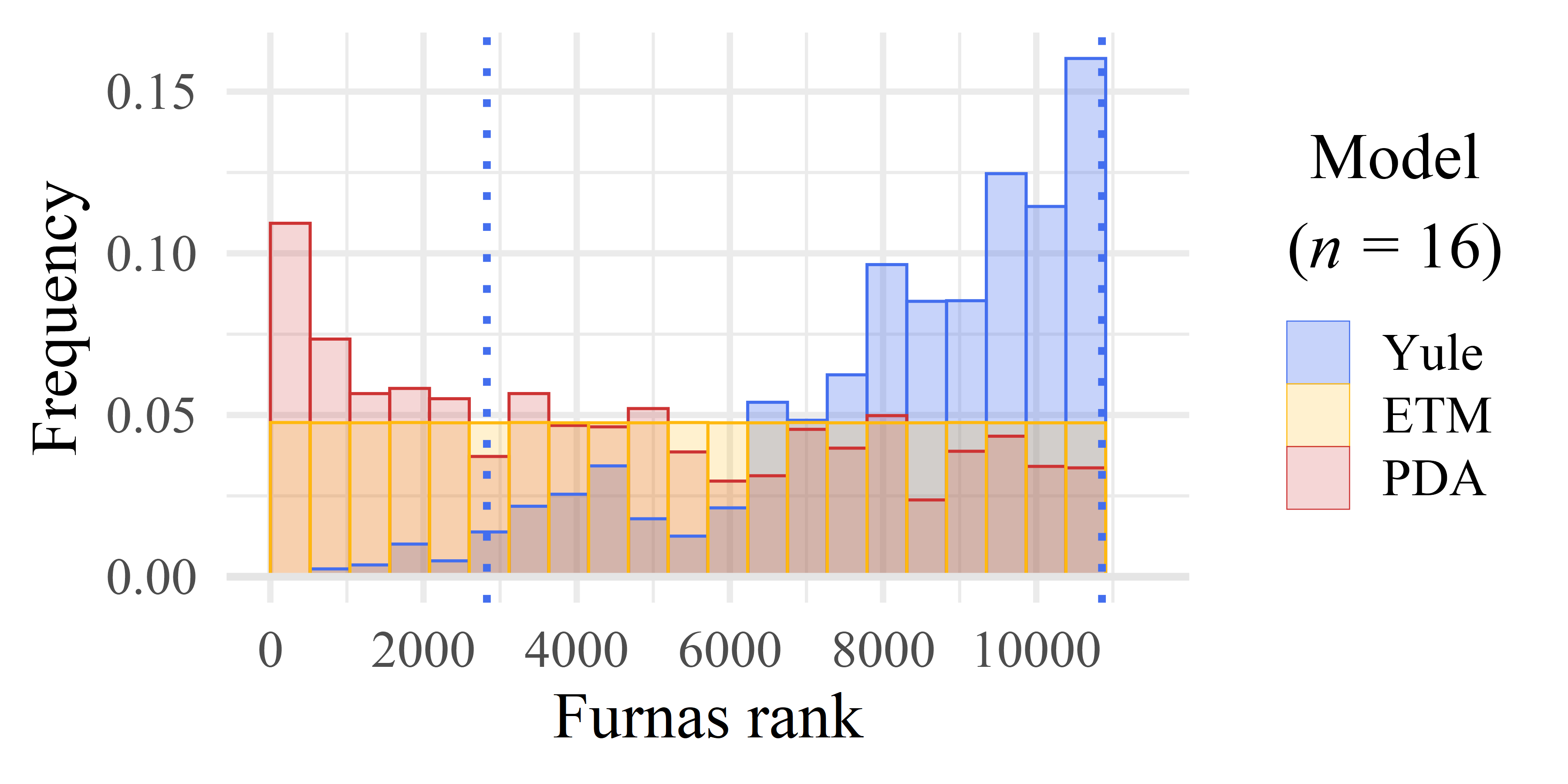}
    \caption{}
    \end{subfigure}
    \begin{subfigure}[t]{0.45\textwidth}
    \centering
    \includegraphics[width=\textwidth]{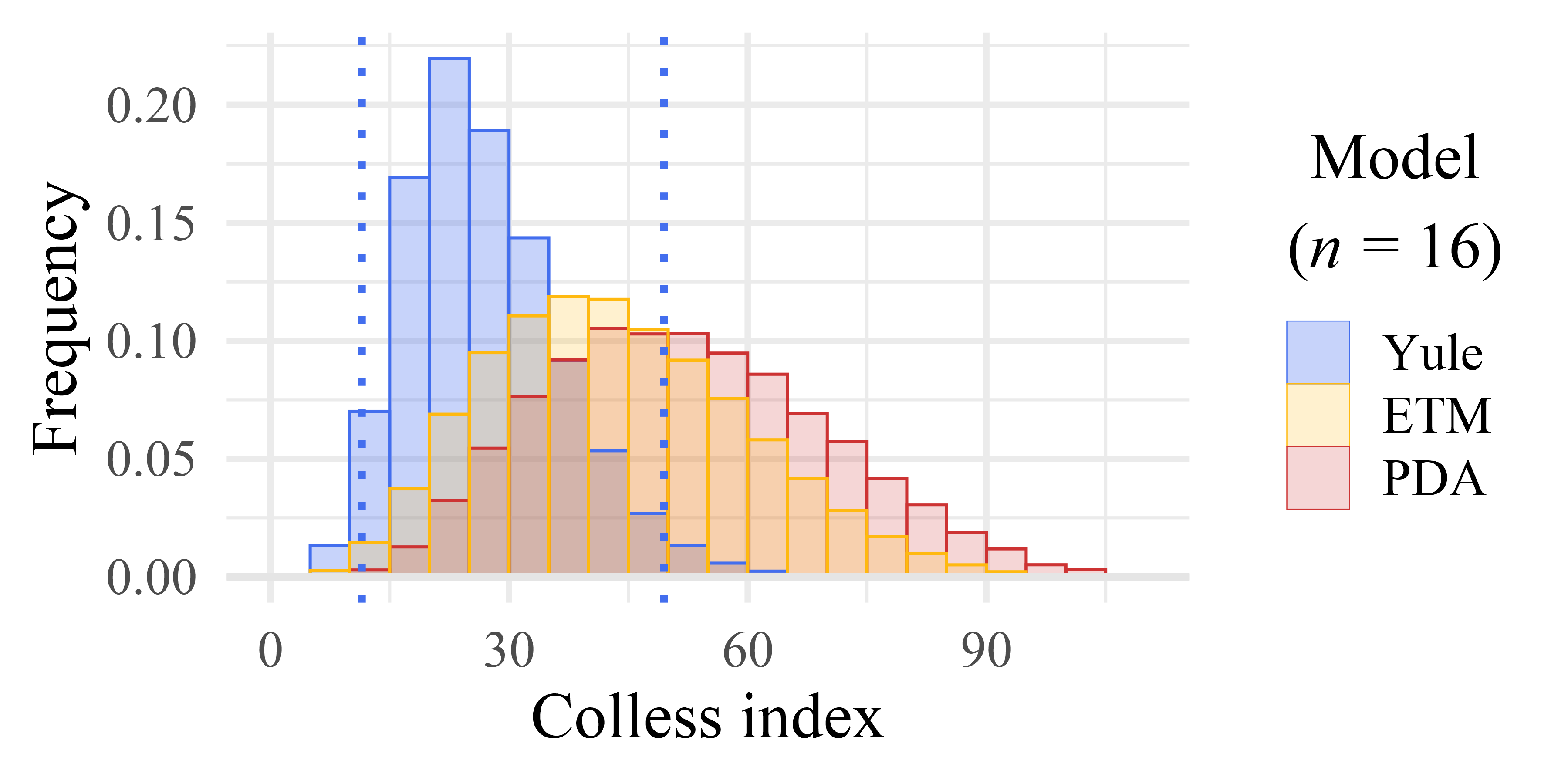}
     \caption{}
    \end{subfigure}\\\vspace{-0.3cm}
    \begin{subfigure}[t]{1\textwidth}
    \centering
    \includegraphics[width=0.96\textwidth]{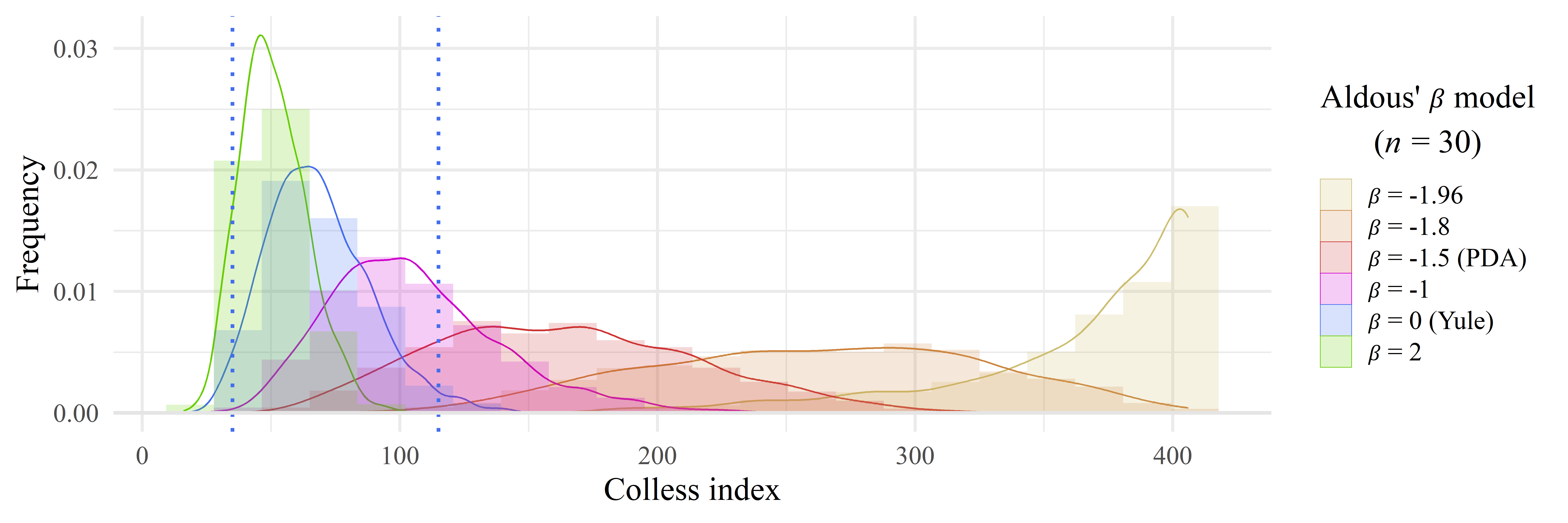}
    \caption{}
    \end{subfigure}\vspace{-0.3cm}
    \caption{Histograms of the exact distributions of (a)  the Furnas rank I-3  and (b) the Colless index I-9 under the Yule model M-0, the PDA model M-15, and the ETM M-16 for $n=16$. The Furnas rank is a balance index, i.e., trees with higher values can be considered as more balanced, and the Colless index is an imbalance index, i.e., trees with higher values can be considered as more imbalanced.
    The uniform distribution of the ETM on $\BTnstar$ is directly visible as the Furnas rank is also a tree enumeration method, which assigns all trees in $\BTnstar$ a unique number in $\{1,...,|\BTnstar|\}$ (here $|\mathcal{BT}^\ast_{16}|=10,905$). The dotted lines indicate the 0.025- and the 0.975-quantile under the Yule model. In Panel (c), histograms of the distributions of the Colless index under the Aldous' $\beta$ splitting model M-1 for several choices of $\beta$ (2,000 sampled trees for each $\beta$ in this case) for $n=30$ are shown with overlaid density curves, estimated using Gaussian kernel density estimation.}
    \label{fig:furnas_distrib}
    \end{figure}

We use two-tailed testing as a foundation for measuring the power of the tree balance indices. In particular, this implies that the alternative hypothesis is \emph{any} deviation from the null model, i.e., that trees are more balanced or imbalanced than expected by chance. The level of confidence $\alpha$ is set to $5\%$, i.e., at most $5\%$ of trees constructed under the null model will be incorrectly assessed as non-$P^0_n$ (\textit{first type error}). Then, given a tree shape statistic $t$, a null model $P^0_n$, an alternative model $P^{a}_n$, and the number of leaves $n$, the testing procedure is as follows:

\begin{enumerate} [(1)]
    \item \textbf{Compute or approximate null distribution of the TSS values:} If $P^0_n$ is either the Yule model M-0, the PDA model M-15, or the ETM M-16, we can compute the exact distribution (function) of $t$ under $P^0_n$ for smaller $n$ (our package \textsf{poweRbal} provides this option for $n\leq 20$) by generating all trees $T\in\BTnstar$ with $n$ leaves, calculating their TSS values and using the probabilities $P_{Y,n}(T)$ given in Equation \eqref{eq:yule_prob} for the Yule model, and $P_{\textup{PDA},n}(T)$ and $P_{\textup{ETM},n}(T)$ given in Table~\ref{tab:mod_distrib} for the PDA and ETM models. \\
    For larger $n$ (or if $P^0_n$ is not Yule, PDA, or ETM) we approximate the distribution (function) by generating $N_d = 10^5$ (except for the PDA and ETM where we use $N_d = 10^4$ to reduce the overall computation time for the many choices of $n$; cf. Figure \ref{fig:comp_pda}) trees under the null model $P^0_n$ and computing their TSS values (see Figure \ref{fig:furnas_distrib}).
    \item \textbf{Create hypothesis test:} The corresponding non-parametric two-tailed test $\Phi$ with small-sample correction, where the null hypothesis that a tree $T\in\BTnstar$ was generated under the null model is rejected if $\Phi(t(T))=1$ and maintained if $\Phi(t(T))= 0$, has the form
    \[\displaystyle \Phi(t(T)) = \begin{cases}
        1 & \text{if } t(T) \not\in [c_1,c_2]\text{, }\\
        0 & \text{if } t(T) \in (c_1,c_2)\text{, }\\
        1 & \text{with probability $p_1$ or $p_2$ if $t(T)=c_1$ or $c_2$, respectively,}\\
        0 &  \text{with probability $(1-p_1)$ or $(1-p_2)$ if $t(T)=c_1$ or $c_2$, respectively,}
    \end{cases}\]
    for a lower and upper bound $c_1, c_2 \in \mathbb{R}$. $c_1$ is chosen as the largest TSS value of step (1) with $P(t(T)<c_1| H_0) \leq \frac{\alpha}{2} = 2.5\%$ and $c_2$ as the smallest TSS value with $P(t(T)>c_2| H_0) \leq \frac{\alpha}{2} = 2.5\%$. Furthermore, we do a small-sample correction \cite[Equation~ 21]{innan_statistical_2005} to ensure that $P(\Phi(t(T))=1|H_0)$ is as close to $\alpha$ as possible (and not lower, which would make the test too conservative). For this purpose, we compute the probabilities $p_1=\frac{\alpha/2 - P(t(T)<c_1| H_0)}{P(t(T)=c_1| H_0)}$ and $p_2=\frac{\alpha/2 - P(t(T)>c_2| H_0)}{P(t(T)=c_2| H_0)}$ to reject the null hypothesis if the TSS value equals the lower or upper bound.
    \item \textbf{Test trees under alternative model:} Construct $N_a=10^3$ trees $(T_i^a)_{i=1,...,N_a}$ under $P^a_n$. For each tree $T_i^a$ compute $t(T_i^a)$ and $\Phi(t(T_i^a))$.
    \item \textbf{Compute power:} Calculate the number $R$ of these trees for which the null model is rejected, i.e, $\Phi(t(T_i^a))=1$. Then, the power of $t$ is the proportion of rejected trees $\frac{R}{N_a}$ or the average over $\Phi^a\coloneq(\Phi(t(T_i^a)))_{i=1,...,N_a}$: \[\text{power}_{(n,P^0_n,P^a_n)}(t)=\overline{\Phi^a}=\frac{1}{N_a}\cdot \sum_{i=1}^{N_a}{\Phi(t(T_i^a))} = \frac{R}{N_a}.\]
    \item \textbf{Compute confidence interval:} As the $\Phi(t(T_i^a))$ are i.i.d. with sample mean $\overline{\Phi^a}$, we can use the central limit theorem and conclude that for large $N_a$ (we use $N_a=10^3$) we approximately have a normal distribution $\overline{\Phi^a} \approx N\left(\mu,\frac{\sigma^2}{N_a}\right)$, where $\mu$ is the overall expected value and $\sigma^2$ the (finite) variance of a random variable $\Phi(t(T_i^a))$. This allows us to estimate the radius of a $(1-\alpha)$\%-confidence interval by approximating $\sigma$ with the sample standard deviation $sd_{\Phi^a}$ (note that $z_\alpha\approx1.96$ for $\alpha=5$\%, i.e., a 95\%-confidence interval): \[\text{radius}_{(n,P^0_n,P^a_n)}(t)= z_\alpha \cdot \frac{\sigma}{\sqrt{N_a}} \approx 1.96 \cdot \frac{\text{sd}_{\Phi^a}}{\sqrt{N_a}} = 1.96 \cdot \frac{\sqrt{\frac{1}{N_a-1}\cdot\sum_{i=1}^{N_a}{(\Phi(t(T_i^a))-\overline{\Phi^a})^2}}}{\sqrt{N_a}} = 1.96 \cdot \frac{\sqrt{\frac{R(N_a-R)}{N_a(N_a-1)}}}{\sqrt{N_a}} .\]
    By using $R(N_a-R)\leq (N_a/2)^2$ as $0\leq R \leq N_a$ we can calculate an upper limit for the radius, namely $1.96 \cdot \sqrt{\frac{1}{4(N_a-1)}}$, which is approximately $0.031\approx3\%$ for $N_a = 10^3$.
\end{enumerate}

\paragraph{Selection of TSS, null model, and alternative models}
In our study, we use all \enquote{non-equivalent} and \enquote{suitable} tree balance indices $t$ listed in Section \myrefSecTBIndices in the supplementary material \cite{FischerKerstingWickeSupp}. We now explain what exactly we mean by this. First, in the present manuscript, we consider tree balance indices as \emph{equivalent} if they induce the same/equivalent statistical test due to differing only in a factor and/or constant (dependent on $n$). All sets of equivalent tree balance indices have, thus, been reduced to their most established representative (marked in bold): $\{$\textbf{Sackin index}, avg. leaf depth, tot. int. path length, tot. path length, avg. vertex depth$\}$, $\{$\textbf{Colless index}, corrected Colless$\}$, and $\{$\textbf{Rogers $J$}, stairs1$\}$. 

Second, we only consider \emph{suitable} indices, which means that we had to exclude one index for practical reasons:  The Colijn-Plazotta ranking produces too high values and thus causes computational problems even for small $n$, like $n=14$, and therefore had to be omitted.

While the cherry index I-22 is not a tree balance index \cite{fischer_tree_2023}, it is included in our study as it has been used in similar contexts already \cite{blum_statistical_2005, matsen_geometric_2006}. The complete list of included TSS can be found in the legends of the graphics.

Furthermore, we use (representatives of) all discrete-time tree models $P_n$ listed in Section \myrefSecTreeModels in the supplementary material \cite{FischerKerstingWickeSupp}. The only exceptions are the most simplistic models that only produce a single tree shape, e.g., the caterpillar or maximally balanced tree (M-17 and M-18). The comparisons are done for $n= 30$ and $100$ to cover some range of tree sizes.

Steps (1)--(5) are performed for all combinations of $n \in \{30,100\}$, TSS, and alternative models.

\paragraph{Software}
In the supplementary material~\cite{FischerKerstingWickeSupp}, we illustrate how to do similar experiments using any set of tree shape statistics, any set of alternative models, and with regard to any null model with the help of our \textsf{R} software package \textsf{poweRbal} provided alongside this manuscript.  While we give more details in the supplementary material, we emphasize that its usage is straightforward. Deciding on a null model, a list of alternative models, and a leaf number $n$, a power analysis encompassing all existing TSS can be done with simple commands (further details on these commands and its options, e.g., specifying the sample sizes, the testing and correction method, and the significance level, can be found in the supplementary material or in the package manual): 
\begin{lstlisting}
    # Get all tss
    tss <- getAllTSS(n)
    # Perform power analysis
    power_data <- powerComp(tss, null_model, alt_models, n)
    # Plot results
    plot(power_data)
\end{lstlisting}

\section{Results} \label{sec:results}
\begin{figure}[ht]
    \centering
    \includegraphics[width=0.98\textwidth]{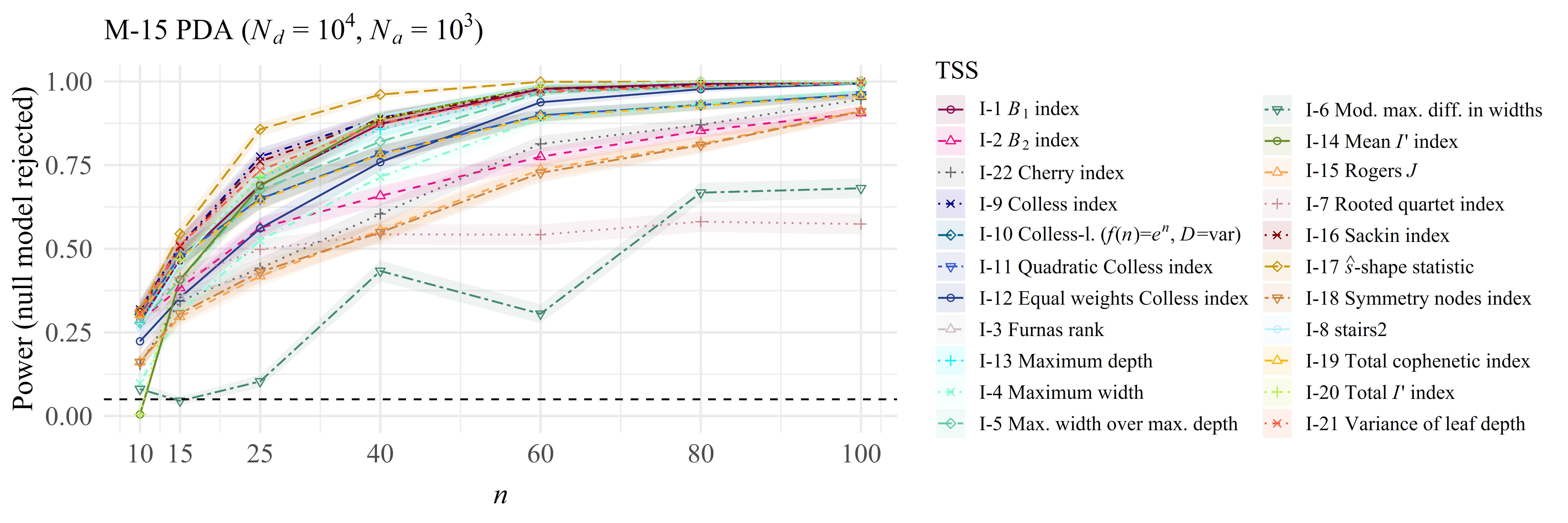}
    \caption{The power of all TSS to correctly identify trees generated under the PDA model M-15 as not having been generated under the Yule model M-0 as a function of $n$. Note that for $n=10$ and $n=15$ the power values are based on the exact TSS distribution under Yule. Notice that the legend of this figure also applies to all subsequent figures.}
    \label{fig:comp_pda}
\end{figure}

\begin{figure}[ht]
    \centering
    \begin{subfigure}[t]{0.49\textwidth}
    \centering
    \includegraphics[width=\textwidth]{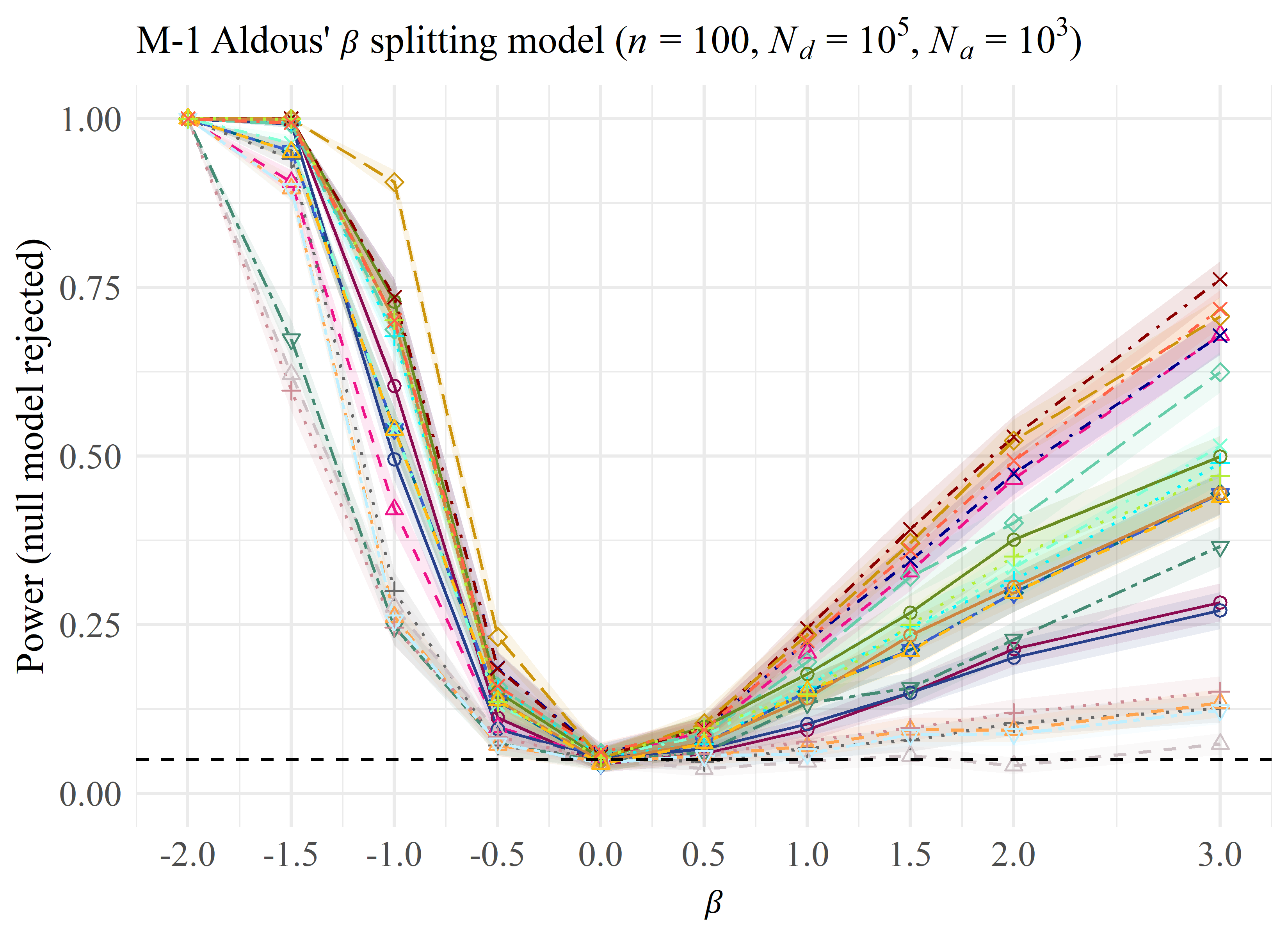}
    \caption{}
    \end{subfigure}
    \begin{subfigure}[t]{0.49\textwidth}
    \centering
    \includegraphics[width=\textwidth]{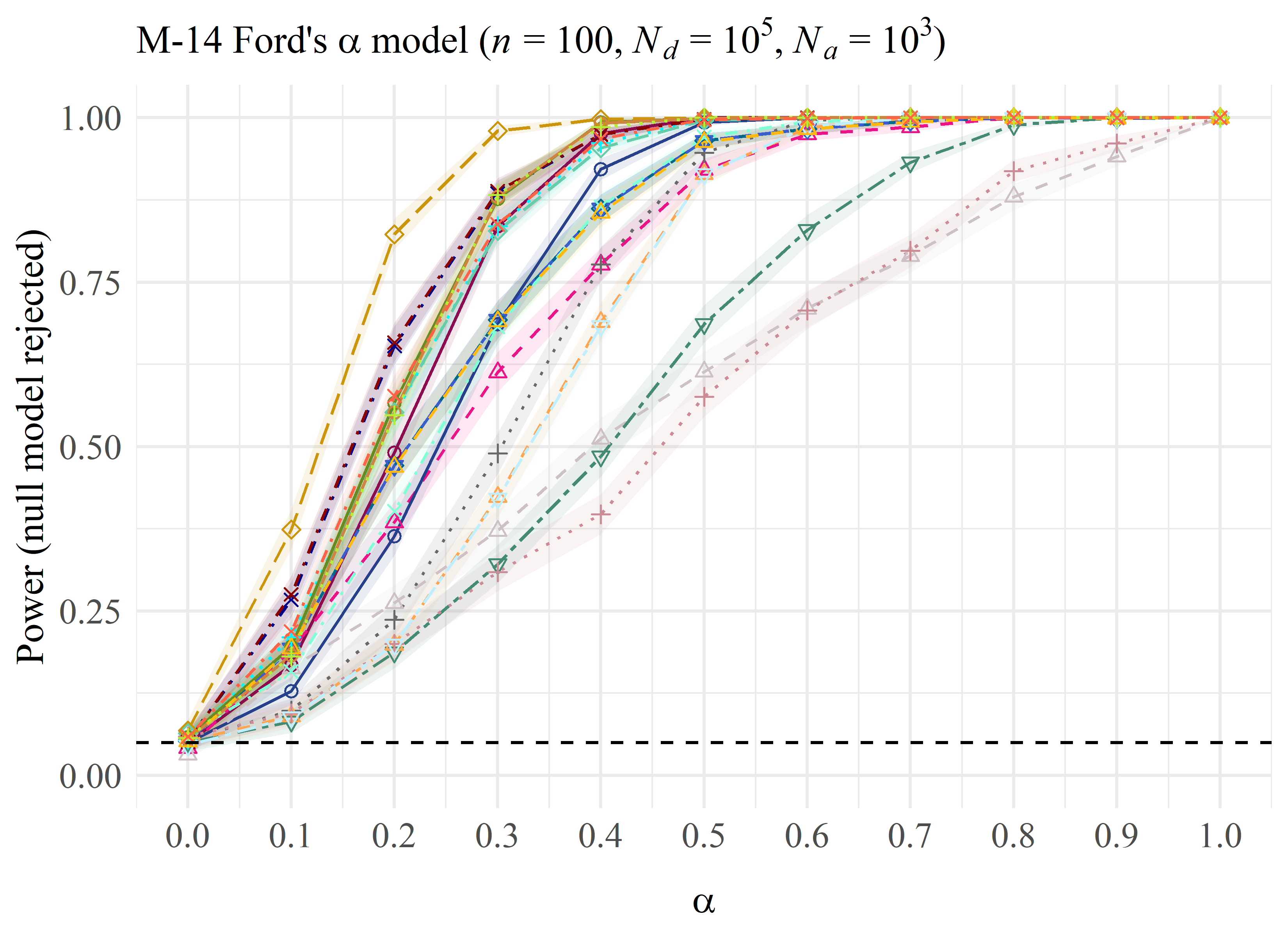}
    \caption{}
    \end{subfigure} \vspace{-0.3cm}
    \caption{The power of all TSS to correctly identify trees generated under (a) the Aldous $\beta$ splitting model M-1 and (b) Ford's $\alpha$ model M-14 (right) as not having been generated under the Yule model M-0 as a function of $\beta$ (resp. $\alpha$). Since for $\beta=0$, Aldous' splitting model corresponds M-1 to the Yule model M-0, all TSS rejected $\approx 0.05$ \% of the trees in that case (as specified with the level of significance). Similarly, since for $\alpha=0$, Ford's $\alpha$ model M-14 corresponds to the Yule model M-0, all TSS rejected $\approx 0.05$ \% of the trees in that case. Notice that this figure follows the same legend as Figure~\ref{fig:comp_pda}.}
    \label{fig:comp_aldous_ford}
\end{figure}

\begin{figure}[ht]
    \centering
    \begin{subfigure}[t]{0.49\textwidth}
    \centering
    \includegraphics[width=\textwidth]{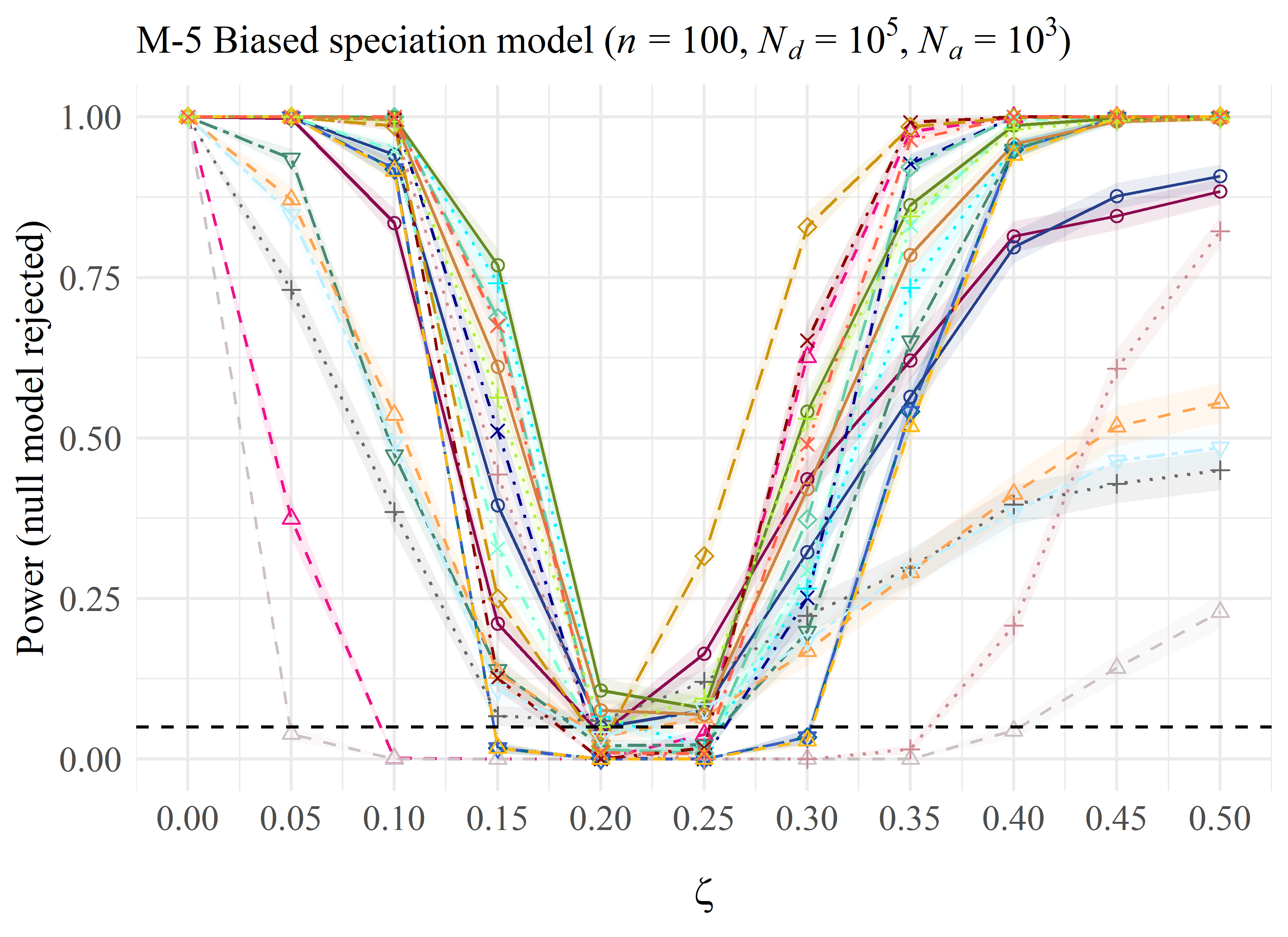}
    \caption{}
    \end{subfigure}
    \begin{subfigure}[t]{0.49\textwidth}
    \centering
    \includegraphics[width=\textwidth]{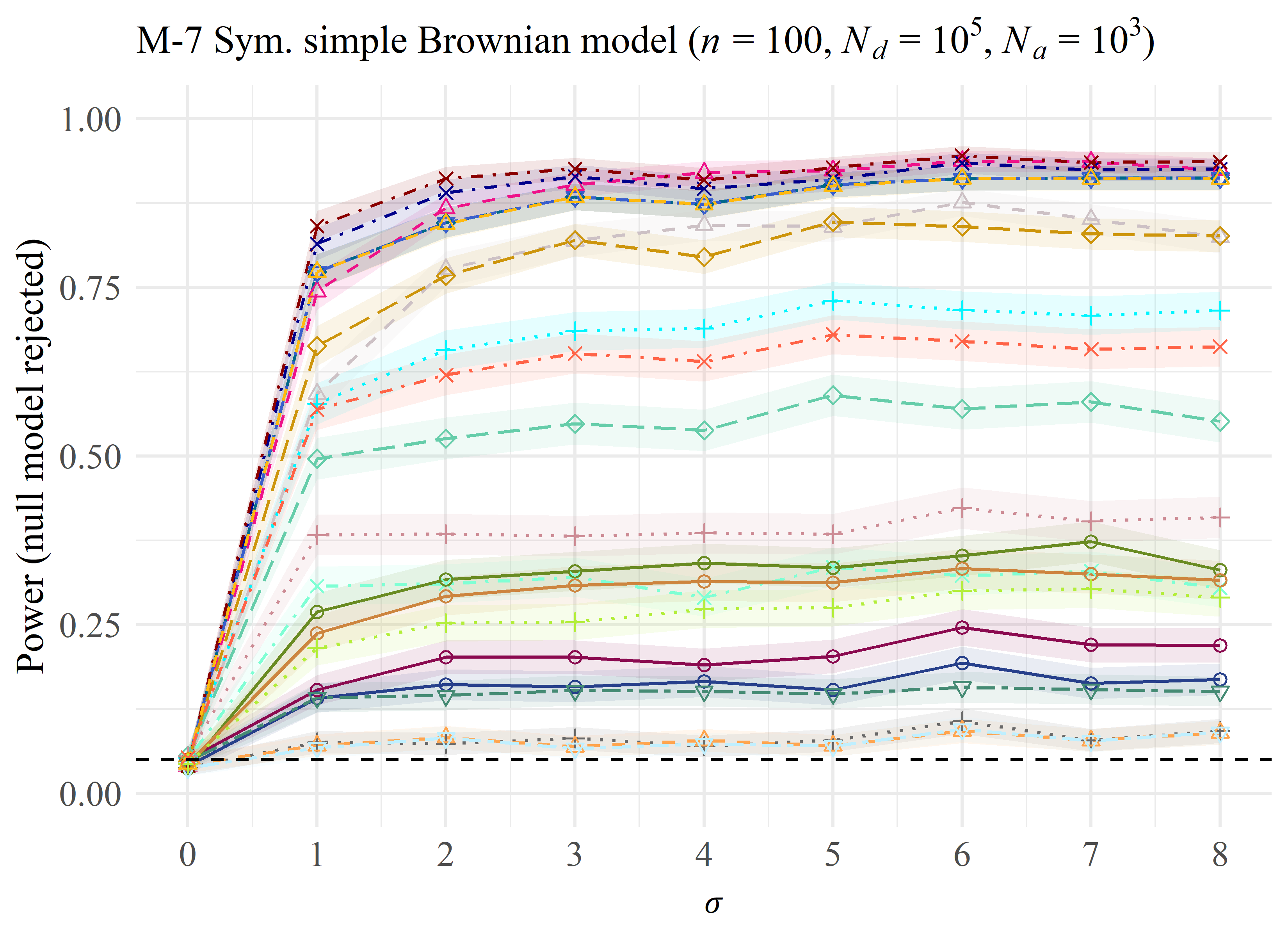}
    \caption{}
    \end{subfigure} \vspace{-0.3cm}
    \caption{The power of all TSS to correctly identify trees generated under (a) the biased speciation model M-5 and (b) the asymmetric simple Brownian model M-7 as not having been generated under the Yule model M-0 as a function of $\zeta$ (resp. $\sigma$). Notice that this figure follows the same legend as Figure~\ref{fig:comp_pda}.}
    \label{fig:comp_biased_simBrown}
\end{figure}

\begin{figure}[ht]
    \centering
    \begin{subfigure}[t]{0.49\textwidth}
    \centering
    \includegraphics[width=\textwidth]{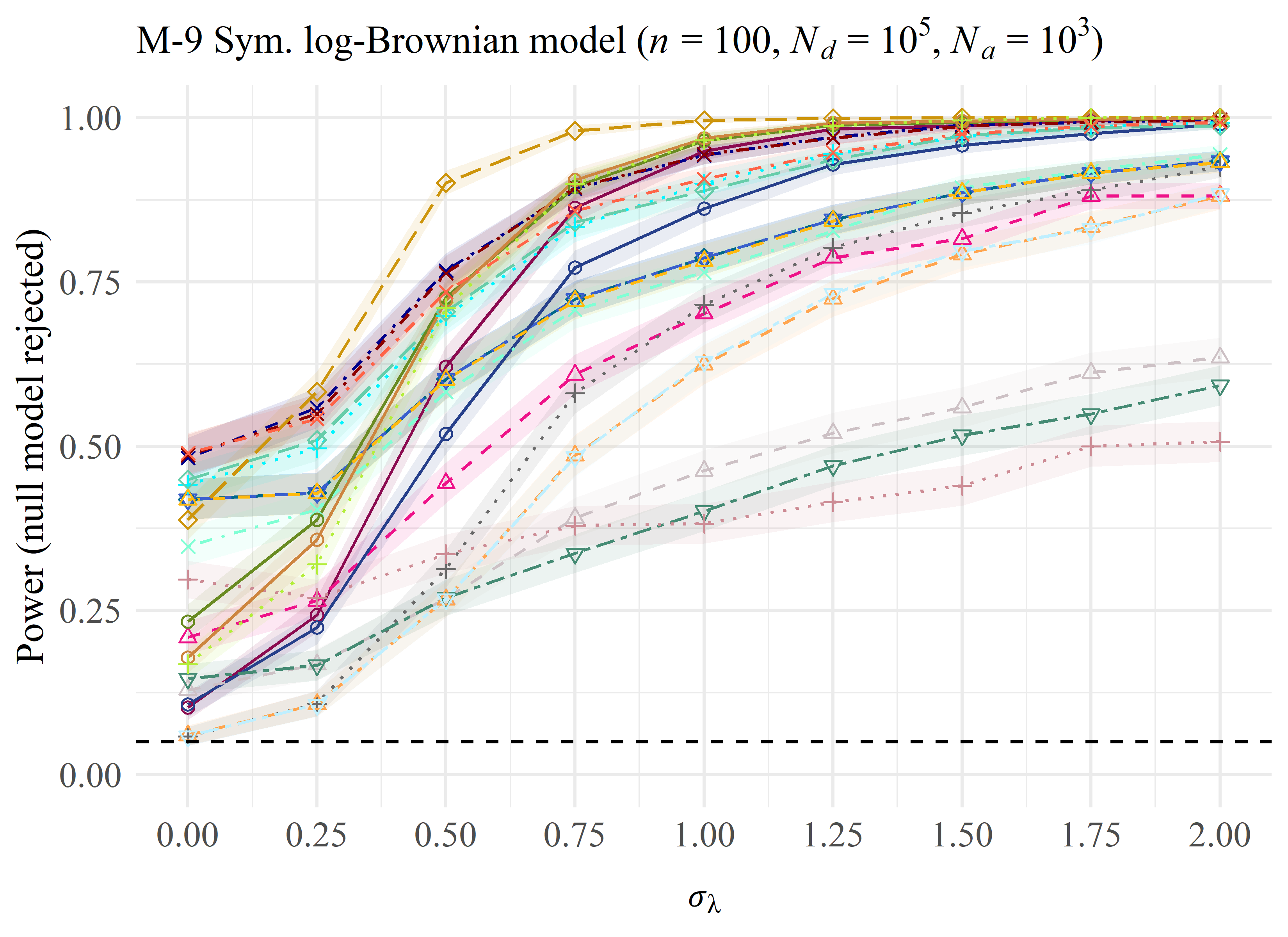}
    \caption{}
    \end{subfigure}
    \begin{subfigure}[t]{0.49\textwidth}
    \centering
    \includegraphics[width=\textwidth]{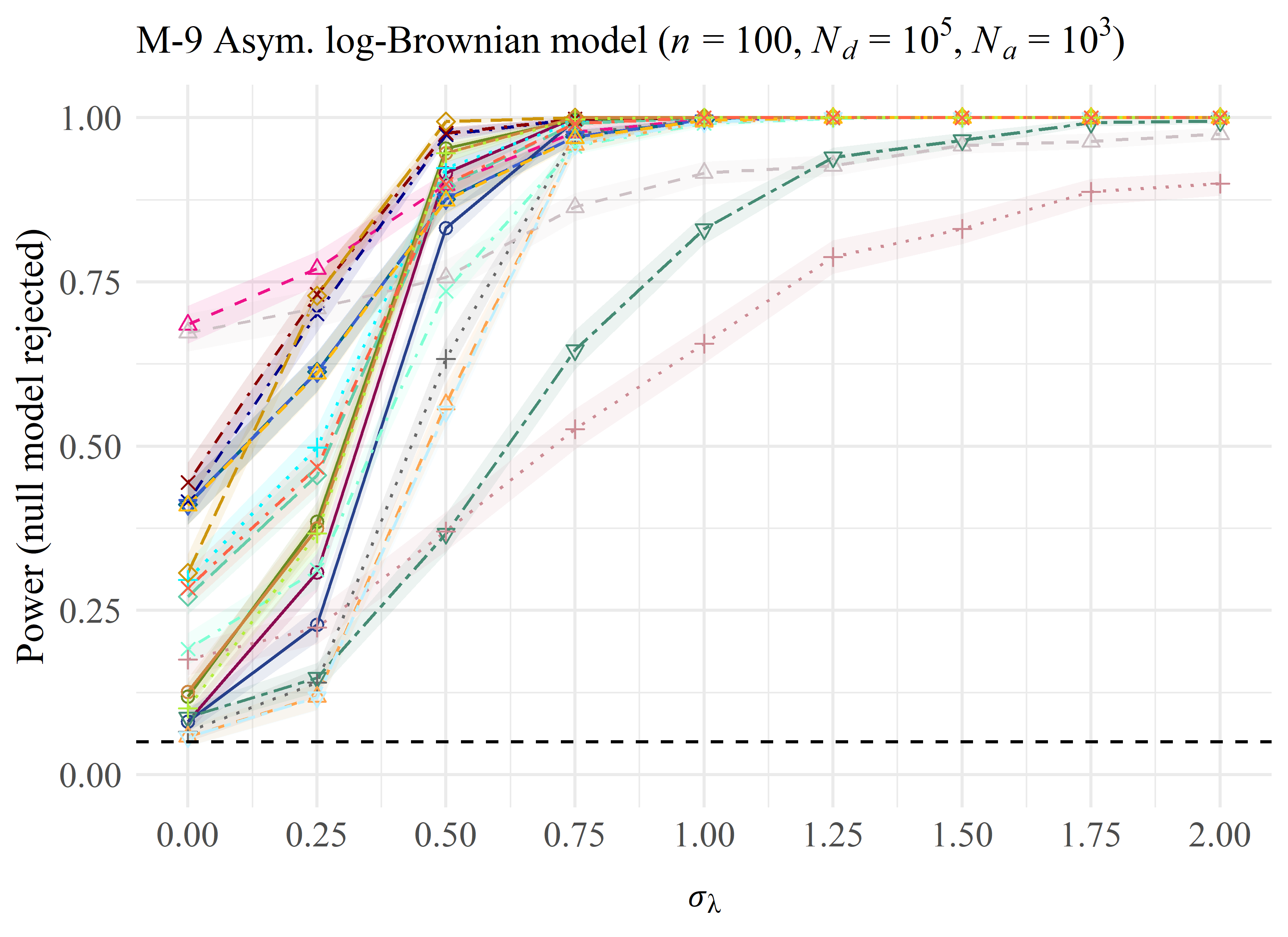}
    \caption{}
    \end{subfigure} \vspace{-0.3cm}
    \caption{The power of all TSS to correctly identify trees generated under (a) the symmetric and (b) the asymmetric punctuated(-intermittent) log-Brownian model M-9 as not having been generated under the Yule model M-0 as a function of $\sigma_\lambda$. The starting trait value is $x_0 = 10$ with constant $\sigma_x=1$. Notice that this figure follows the same legend as Figure~\ref{fig:comp_pda}.}
    \label{fig:comp_logBrown}
\end{figure}

\begin{figure}[ht]
    \centering
    \begin{subfigure}[t]{0.49\textwidth}
    \centering
    \includegraphics[width=\textwidth]{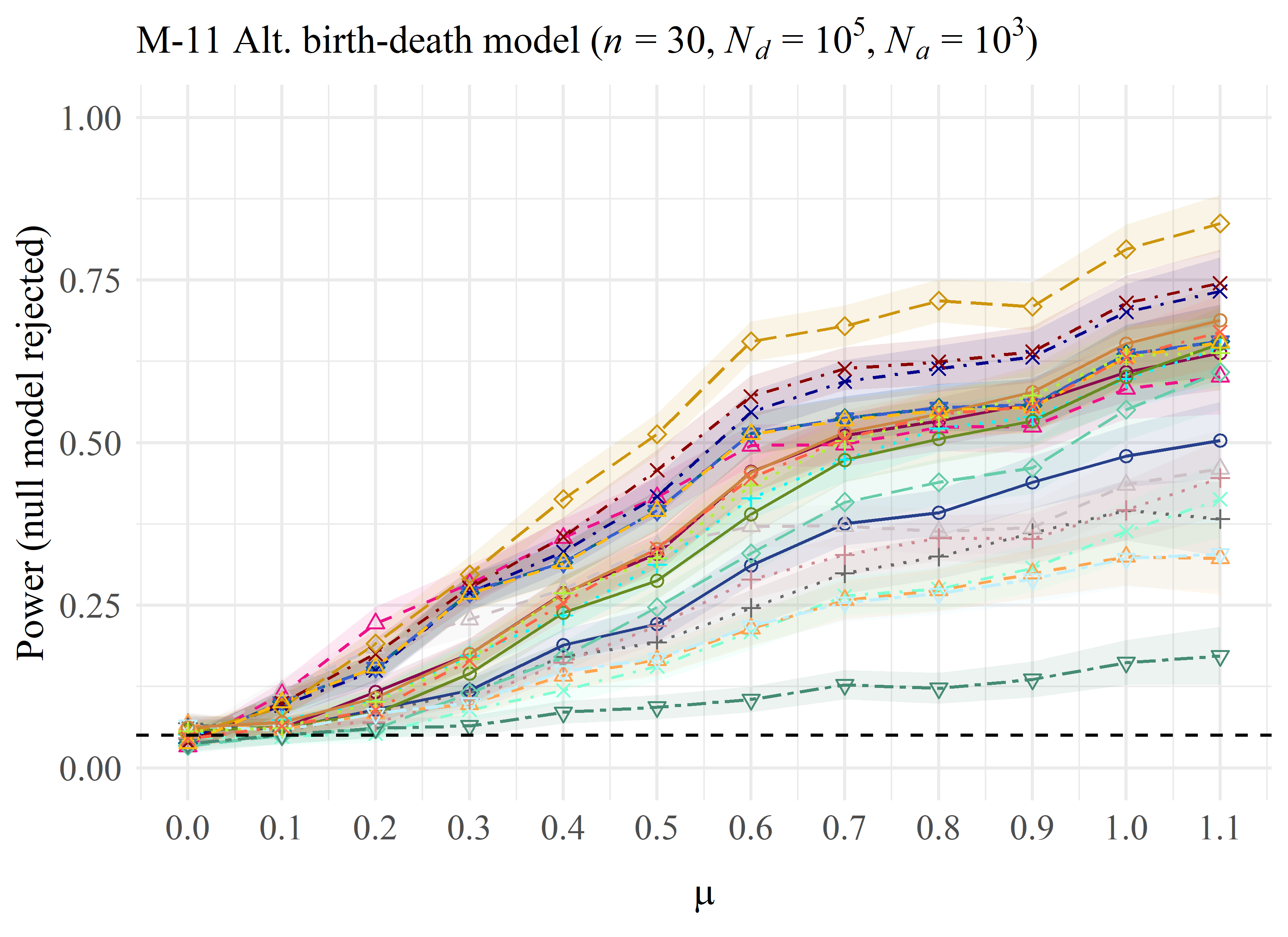}
    \caption{}
    \end{subfigure}
    \begin{subfigure}[t]{0.49\textwidth}
    \centering
    \includegraphics[width=\textwidth]{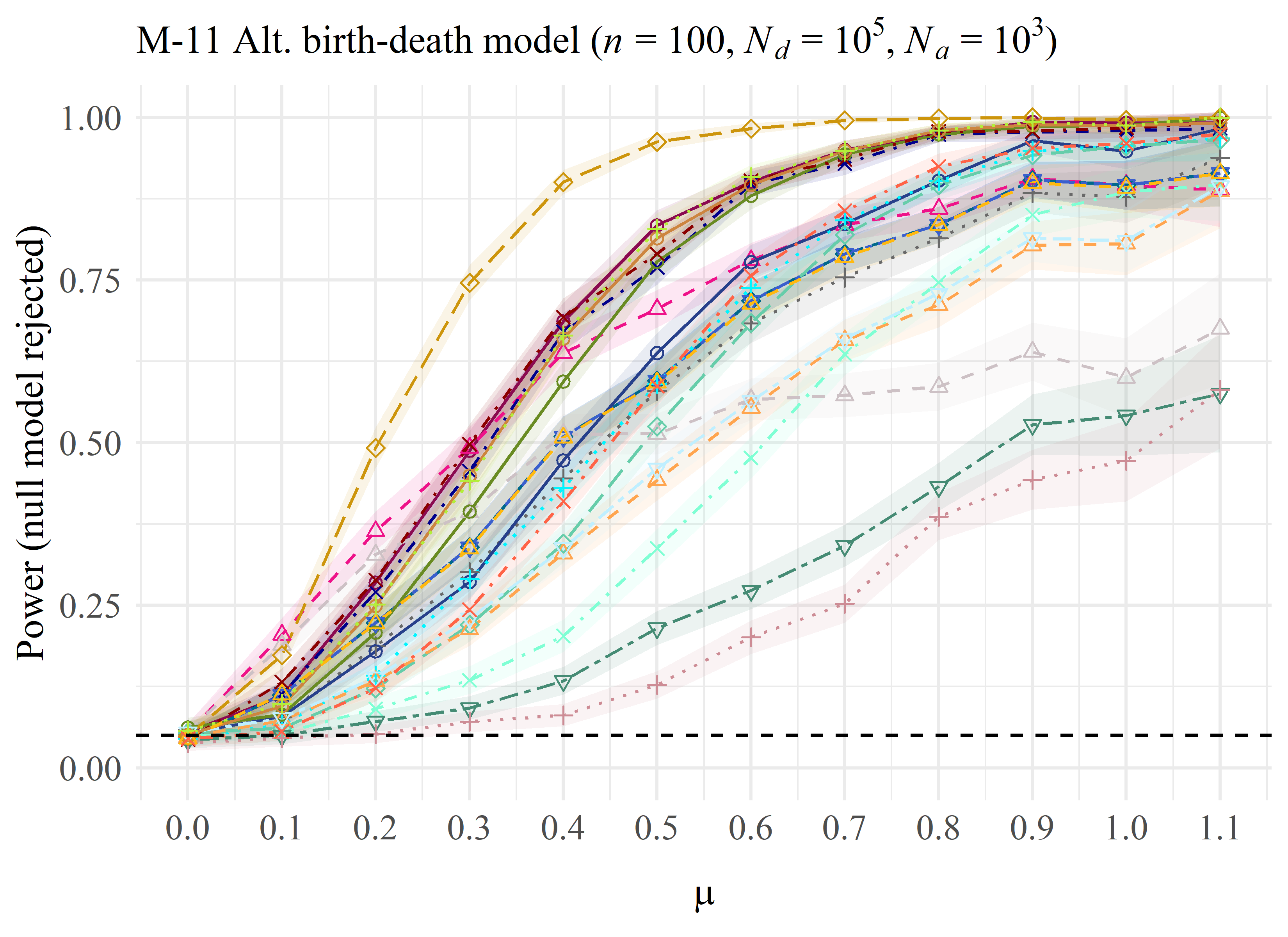}
    \caption{}
    \end{subfigure} \vspace{-0.3cm}
    \caption{The power of all TSS to correctly identify trees generated under the alternative birth-death model M-11 as not having been generated under the Yule model M-0 as a function of $\mu$ for two different tree sizes ((a) $n=30$ and (b) $n=100$). Notice that this figure follows the same legend as Figure~\ref{fig:comp_pda}.}
    \label{fig:comp_altbirthdeath}
\end{figure}

In this section, we summarize the most important results obtained in our simulation study. Figures~\ref{fig:comp_pda}--\ref{fig:comp_altbirthdeath} show the power (with 95\%-confidence bands) of the various TSS to correctly identify trees generated under models distinct from the Yule model M-0 as not having been generated under the Yule model. The power is plotted either against the leaf number $n$ (Figure~\ref{fig:comp_pda}) or against a tree model parameter (Figures~\ref{fig:comp_aldous_ford}--\ref{fig:comp_altbirthdeath}). Note that all figures follow the legend of Figure~\ref{fig:comp_pda}.  Further result figures can be found in Section \myrefSecPowerResults of the supplementary material \cite{FischerKerstingWickeSupp}. As evident from all our figures, the curves look quite different for different alternative models. In the following, we study some aspects of these figures and conclusions to be drawn more in-depth.

\paragraph{The most and least powerful TSS} A central goal of our study was to identify TSS that exhibit a high power for discriminating between different (macro)evolutionary  models. Our results show that the performance of different TSS clearly depends on the alternative models and also on the null model used (recall that we use the Yule model M-0 as the null model throughout the present manuscript, but in Section~\myrefOtherNullModels of the supplementary material \cite{FischerKerstingWickeSupp} we also briefly discuss other null models).  Even for a fixed alternative model, there are sometimes striking differences for different parameter choices. 
For example, considering the log-Brownian models M-9 in Figure~\ref{fig:comp_logBrown} (a) \& (b), we can observe that the $B_1$ index~I-1 and the equal weights Colless index I-12 are among the TSS with the lowest power for small values of $\sigma_\lambda$, whereas they are among the most powerful TSS for larger values of $\sigma_\lambda$.
A related observation is that in some cases, there are few to no crossings between the power curves for different TSS (e.g., Figures~\ref{fig:comp_aldous_ford} (a) and \ref{fig:comp_biased_simBrown} (b)), implying that the ranking of the TSS by their power does not change across the different settings tested, whereas for others, there are significant changes (e.g., Figures~\ref{fig:comp_biased_simBrown} (a) and~\ref{fig:comp_logBrown} (a) \& (b)).

Nevertheless,  we can observe that there are some TSS that (almost) consistently outperform the others, such as the $\widehat{s}$-shape statistic I-17, whereas others, such as the modified maximum difference in widths I-6 or the rooted quartet index I-7, tend to consistently exhibit a low power. In all cases, there are notable exceptions, though. In the case of the $\widehat{s}$-shape statistic I-17 it is interesting to note that it is outperformed by other TSS (such as the $B_2$ index I-2 or the Sackin index I-16) in the symmetric simple Brownian model M-7 (Figure~\ref{fig:comp_biased_simBrown} (b)) as well as for parameter-model combinations close to the Yule model M-0 in the asymmetric log-Brownian model M-9  (Figure~\ref{fig:comp_logBrown} (b))). In the case of the modified maximum difference in widths I-6, exceptions to its overall relatively low power are for instance Aldous' $\beta$ splitting model M-1 (Figure~\ref{fig:comp_aldous_ford} (a)) and the biased speciation model M-5 (Figure~\ref{fig:comp_biased_simBrown} (a)). Another exception for the rooted quartet index I-7 is the simple Brownian model M-7 (Figure~\ref{fig:comp_biased_simBrown} (b)). It is also noticeable that while the power of TSS should increase with increasing $n$, and this is indeed the case for almost all TSS (see Figure~ \ref{fig:comp_pda} and the differences between $n=30$ and $n=100$ in Figure~\ref{fig:comp_altbirthdeath} (a) \& (b)), the power of the modified maximum difference in widths I-6 does not seem to follow this pattern for the PDA model M-15 (Figure~\ref{fig:comp_pda}).

Finally, we remark that it is often the case that several TSS have equal power, indicated by their curves being clumped together, as is for instance the case for the top curves in the simple Brownian model M-7 (Figure~\ref{fig:comp_biased_simBrown} (b)). In other cases, the curves are clearly separated, as for instance in the biased speciation model M-5 (Figure~\ref{fig:comp_biased_simBrown} (a)). Interestingly, the total cophenetic index I-19 and the quadratic Colless index I-11 show extremely similar behavior (their lines are nearly always right on top of each other), and it would be interesting to investigate this further. In summary, we thus reiterate that which TSS have a high power and which have a low power depends on the model setting, highlighting the importance of performing a power analysis such as the one presented here prior to downstream analyses.

\paragraph{Fitting tree models} For parameter settings that correspond to the Yule model M-0, the power of all TSS should be close to 0.05 and this is clearly visible in the corresponding figures (e.g., for $\beta =0$ or $\alpha=0$ in Figure~\ref{fig:comp_aldous_ford} (a) \& (b), or $\mu = 0$ in Figure~\ref{fig:comp_altbirthdeath} (a) \& (b)). Interestingly, we also observe a power close to 0.05 for almost all TSS for the biased speciation model M-5 for $\zeta$ between 0.20 and 0.25 (Figure~\ref{fig:comp_biased_simBrown} (a)). This could indicate that the biased speciation model M-5 is very close to or coincides with the Yule model M-0 for $\zeta$ in this range. However, it could also mean that none of the TSS employed in this study can detect deviation from the Yule model M-0 in this setting, even though the models might theoretically not be identical. This leads us to the following observation for fitting tree models to empirical data: When a collection of empirical trees is tested against a specific evolutionary model and a low power is observed for all TSS, this can be an indication of a good fit of the model to the data. However, low power is only a sufficient and not a necessary condition for a good model fit, as it could be the case that other TSS are needed to detect differences between the data and the model. Thus, caution is warranted. A good example illustrating this can be found in Figure \myrefFigABDvsAldous in the supplementary material \cite{FischerKerstingWickeSupp}, where we use Aldous' $\beta$ splitting model M-1 with $\beta = -1$ as the null model and the alternative birth-death model as the alternative model. In this case, almost none of the TSS can tell the alternative birth-death trees with parameter $\mu=0.4$ apart from the Aldous $\beta=-1$ trees, except for the $B_2$ index I-2 and the Furnas rank I-3 (where the latter is even more surprising as it is primarily used as a tree enumeration method and not often employed in applications). Without including these two indices, one might have concluded that both models are equivalent.

\paragraph{Influence of the speciation mode} Finally, we remark that next to specific parameter choices, there are other factors that can impact the power of the different TSS. We illustrate this using the example of \emph{symmetric} versus \emph{asymmetric speciation} (also known as one- or two-daughter-change \cite{heard_patternsrate_1996}) for rate-varying models: Symmetric speciation means that after speciation two identical \enquote{child} species (of age 0) are created, while asymmetric speciation means that the \enquote{parent} species persists (and keeps its current age), and only one new \enquote{child} species (of age 0) is created. Models that allow the usage of both symmetric and asymmetric speciation are marked with $\SymAsym$ in Table~\ref{tab:mod_growing_TS} as well as in the tables in the supplementary material. In Figure~\ref{fig:comp_logBrown} (a) \& (b), we depict the power curves for a symmetric and an asymmetric log-Brownian model M-9 with otherwise identical parameters (i.e., the only difference between the two plots is the speciation mode) and there are some striking differences. For instance, the $B_2$ index I-2 is among the most powerful indices in the asymmetric case, but its power decreases in the symmetric case. In the asymmetric case, all TSS also reach a high power sooner than in the symmetric case, indicating that deviation from the Yule model M-0 is easier to detect assuming asymmetric speciation. The second observation is perhaps unsurprising since the Yule model M-0 is itself a symmetric model. Nevertheless, we can conclude that seemingly subtle differences between different models such as the speciation mode, may have a significant impact on the power of TSS.

\section{Conclusion and outlook} \label{sec:conclusion}

Tree shape statistics in general and tree balance indices in particular are powerful tools to find realistic models for the evolution of sets of species under investigation. However, not all these statistics are equally suitable for detecting every evolutionary model.

In most of our analyses, the $\widehat{s}$-shape statistic I-17 was the most powerful statistic discriminating alternative models from the Yule model M-0. While this nicely complements a study by Blum at al. \cite{blum_which_2006}, who found that the $\widehat{s}$-shape statistic I-17 \enquote{warrants maximal power for rejecting the [Yule] model against the PDA [model]}, this is hardly surprising. It can easily be explained by the fact that the $\widehat{s}$-shape statistic I-17 was even introduced with that very purpose in mind. In fact, except for the normalizing constant of $P_{Y,n}$, the $\widehat{s}$-shape statistic I-17 corresponds to the logarithm of $P_{Y,n}$ (c.f. \cite{semple_phylogenetics_2003}), i.e., of the probability of generating a tree under the Yule model M-0. However, while Blum et al. in their study explicitly did not claim that the $\widehat{s}$-shape statistic I-17 is \enquote{generally superior}, in our study we could see that it is indeed superior concerning many models, not only the PDA model M-15. Some exceptions include the Brownian models M-7 and M-9, in which the $\widehat{s}$-shape statistic I-17 is outperformed by other TSS, cf. Figures \ref{fig:comp_biased_simBrown} (b) and \ref{fig:comp_logBrown} (b). Khurana et al. ~\cite{khurana_limits_2023} also found that for their data, the $\widehat{s}$-shape statistic I-17 had relatively low power, whereas the $B_2$ index I-2, the variance of leaf depths I-21, the mean and total $I$ indices I-14 and I-20 (without correction method $I'$), stairs2 I-8, and the normalized (corrected) Colless index (I-9) were most powerful in discriminating empirical trees from trees simulated under a constant-rate-birth-death model (which induces the same distribution on $\BTnstar$ as the Yule model M-0). 
In Section \myrefOtherNullModels of the supplementary material we discuss their findings further and show how to approach their research questions with our methods.
We also remark that the findings by Khurana et al. ~\cite{khurana_limits_2023} concerning the $B_2$ index I-2 nicely complement a study by Bienvenu et al. \cite{bienvenu_revisiting_2020}, which used five different null models (one of which was the Yule model M-0) and tested six different balance indices for their power (note that they did not include the $\widehat{s}$-shape statistic I-17 in their study, though). In this study, the authors found that the $B_2$-index I-2 has on average higher power than the other investigated indices.
Note that in most of our studies, too, the $B_2$ index I-2 was amongst the most powerful indices.

As shown by Bienvenu et al.~\cite{bienvenu_revisiting_2020}, the choice of different null models can make a huge difference. However, as our results show, even with the Yule model M-0 as a null model, there are huge performance differences between the different TSS. These differences might imply that there are different patterns of imbalance produced by the different models, some of which can only be recognized by a subset of tree shape statistics. Therefore, the choice of the most powerful TSS for each individual case is important. We thus highly recommend to perform a power analysis such as the one presented here prior to downstream analyses. For this purpose, our software package \textsf{poweRbal} is specifically designed in such a way that it also caters to researchers unfamiliar with the programming language \textsf{R}. Trees generated by other software tools can easily be loaded into our package (e.g., in the well-known Newick format) and used as a basis for an alternative model. Newly developed indices and models can easily be added to the package, so that the user can test several variations to see which one is most powerful in the given setting. \textsf{poweRbal} also enables researchers to perform a simulation study beforehand to decide on which TSS to use and to subsequently analyze their data with the respective optimal tools.

Concerning theoretical research, on the other hand, a goal for future developments could be to  develop tools to guess the most likely model simply based on a given tree. First results in this regard were presented in \cite{kaur_distributions_2023}, where the likelihood of pitchforks and cherries under Ford's $\alpha$-model M-14 has been calculated, which enables users to find the most likely $\alpha$-value. However, as the possibilities of using pitchforks and cherries for estimating the best-fitting overall model are limited, more research in this regard is necessary. 

Another interesting direction for future research would be to investigate whether there is a connection between the power and resolution of different TSS. Specifically, it is interesting to note that Fischer et al.~\cite[Chapter~26]{fischer_tree_2021} recently observed that the $\widehat{s}$-shape statistic I-17  had a high resolution for two empirical data sets (meaning that it showed few ties), whereas the modified maximum difference in widths I-6, an index we observed to have a relatively low power in a variety of settings, showed a low resolution (meaning that it showed many ties). This could indicate a connection between resolution and power for some but not all TSS. For the rooted quartet index I-7, for instance, Fischer et al.~\cite[Chapter~26]{fischer_tree_2021} observed a high resolution while we observe a relatively low power of this index in the present analysis. Nevertheless, exploring why certain TSS behave in distinct ways and whether there is a connection  between power and resolution for some TSS is an interesting direction for future research.

Finally, our manuscript has focused on basic models, i.e., on models which contain clear instructions on how to generate (phylogenetic) trees. However, in the literature also more elaborate and complex models can be found (e.g., \cite[pp. 9--13]{mooers_models_2007} and \cite{marquitti_allopatry_2020}). For instance, evolution can be modeled at the genome level, so that subsequently a phylogenetic tree can be reconstructed (e.g., \cite{marquitti_allopatry_2020}). For the latter, different reconstruction methods can be used, and it has been shown that these methods also have an impact on the resulting tree's balance (see, e.g., \cite{rohlf_accuracy_1990,huelsenbeck_phylogenetic_1996}, \cite[p.~1141]{blum_matrilineal_2006}). Moreover, we have primarily focused on models from phylogenetics, but tree models also frequently occur in other areas such as population genetics, probability theory, or computer science. While some of the models used in these areas are closely related to the Yule model M-0 (see, e.g., \cite{Fuchs2024} and Remark~\myrefRemarkPopgen in the supplement \cite{FischerKerstingWickeSupp} of the present manuscript), it would be an interesting direction of future research to include models such as Aldous' continuum random trees ~\cite{Aldous_1991_ContI,Aldous_1991_ContII,Aldous_1993_ContIII} in the analyses.

\paragraph{Acknowledgement} MF and SK were supported by the project ArtIGROW, which is a part of the WIR!-Alliance “ArtIFARM – Artificial Intelligence in Farming”, and gratefully acknowledge the Bundesministerium für Bildung und Forschung (Federal Ministry of Education and Research, FKZ: 03WIR4805) for financial support. Moreover, we wish to thank Volkmar Liebscher for helpful advice on the development of the \textsf{R} package and Annalena Franke for implementing first versions of several tree models. Additionally, we thank three reviewers and the editor Noah Rosenberg for detailed comments on an earlier version of this paper.

\vskip2pc

\bibliographystyle{RS} 
\bibliography{literature} 

\begin{thebibliography}{99}

\bibitem{Futuyma_evolution_2001}
Futuyma DJ, Meagher TR. 2001  Evolution, science and society: {E}volutionary
  biology and the national research agenda. {\em California Journal of Science
  Education} \textbf{1}, 19--32.

\bibitem{raup_stochastic_1973}
Raup DM, Gould SJ, Schopf TJM, Simberloff DS. 1973  Stochastic Models of
  Phylogeny and the Evolution of Diversity. {\em The Journal of Geology}
  \textbf{81}, 525--542.
(\href{http://dx.doi.org/10.1086/627905}{10.1086/627905})

\bibitem{Raup_stochastic_1974}
Raup DM, Gould SJ. 1974  Stochastic Simulation and Evolution of
  Morphology-Towards a Nomothetic Paleontology. {\em Systematic Biology}
  \textbf{23}, 305--322.
(\href{http://dx.doi.org/10.1093/sysbio/23.3.305}{10.1093/sysbio/23.3.305})

\bibitem{Schopf_genomic_1975}
Schopf TJM, Raup DM, Gould SJ, Simberloff DS. 1975  Genomic versus morphologic
  rates of evolution: influence of morphologic complexity. {\em Paleobiology}
  \textbf{1}, 63--70.
(\href{http://dx.doi.org/10.1017/s0094837300002207}{10.1017/s0094837300002207})

\bibitem{Gould_shape_1977}
Gould SJ, Raup DM, Sepkoski JJ, Schopf TJM, Simberloff DS. 1977  The shape of
  evolution: a comparison of real and random clades. {\em Paleobiology}
  \textbf{3}, 23--40.
(\href{http://dx.doi.org/10.1017/s009483730000508x}{10.1017/s009483730000508x})

\bibitem{Raup_stochastic_1977}
Raup DM. 1977  Chapter 3 {S}tochastic Models in Evolutionary Palaeontology. In
  {\em Developments in Palaeontology and Stratigraphy} ,  pp. 59--78. Elsevier.
(\href{http://dx.doi.org/10.1016/s0920-5446(08)70323-6}{10.1016/s0920-5446(08)70323-6})

\bibitem{Schopf_evolving_1979}
Schopf TJM. 1979  Evolving paleontological views on deterministic and
  stochastic approaches. {\em Paleobiology} \textbf{5}, 337--352.
(\href{http://dx.doi.org/10.1017/s0094837300006606}{10.1017/s0094837300006606})

\bibitem{edwards_reconstruction_1964}
Edwards AWF, Cavalli-Sforza LL. 1964 p. 17–27.
In {\em Reconstruction of evolutionary trees}, p. 17–27. Cambridge University
  Press.
(\href{http://dx.doi.org/10.1017/9781316276259.003}{10.1017/9781316276259.003})

\bibitem{edwards_estimation_1970}
Edwards AWF. 1970  Estimation of the Branch Points of a Branching Diffusion
  Process. {\em Journal of the Royal Statistical Society Series B: Statistical
  Methodology} \textbf{32}, 155–164.
(\href{http://dx.doi.org/10.1111/j.2517-6161.1970.tb00828.x}{10.1111/j.2517-6161.1970.tb00828.x})

\bibitem{norstrom_phylotempo_2012}
Norstr{\"o}m MM, Prosperi MCF, Gray RR, Karlsson AC, Salemi M. 2012
  {PhyloTempo}: a set of {R} scripts for assessing and visualizing temporal
  clustering in genealogies inferred from serially sampled viral sequences.
  {\em Evolutionary Bioinformatics} \textbf{8}, 261--269.
(\href{http://dx.doi.org/10.4137/EBO.S9738}{10.4137/EBO.S9738})

\bibitem{colijn_phylogenetic_2014}
Colijn C, Gardy J. 2014  Phylogenetic tree shapes resolve disease transmission
  patterns. {\em Evolution, Medicine, and Public Health} \textbf{2014},
  96--108.
(\href{http://dx.doi.org/10.1093/emph/eou018}{10.1093/emph/eou018})

\bibitem{pompei2012phylogenetic}
Pompei S, Loreto V, Tria F. 2012  Phylogenetic Properties of {RNA} Viruses.
  {\em PLoS ONE} \textbf{7}, e44849.
(\href{http://dx.doi.org/10.1371/journal.pone.0044849}{10.1371/journal.pone.0044849})

\bibitem{holman_languages_2010}
Holman EW. 2010  Do languages originate and become extinct at constant rates?.
  {\em Diachronica} \textbf{27}, 214--225.
(\href{http://dx.doi.org/10.1075/dia.27.2.03hol}{10.1075/dia.27.2.03hol})

\bibitem{fischer_tree_2023}
Fischer M, Herbst L, Kersting SJ, K{\"u}hn L, Wicke K. 2023 {\em Tree Balance
  Indices - A Comprehensive Survey}.
Berlin: Springer.

\bibitem{farris_expected_1976}
Farris JS. 1976  Expected asymmetry of phylogenetic trees. {\em Systematic
  Zoology} \textbf{25}, 196.
(\href{http://dx.doi.org/10.2307/2412748}{10.2307/2412748})

\bibitem{harding_probabilities_1971}
Harding EF. 1971  The probabilities of rooted tree-shapes generated by random
  bifurcation. {\em Advances in Applied Probability} \textbf{3}, 44--77.
(\href{http://dx.doi.org/10.2307/1426329}{10.2307/1426329})

\bibitem{rosen_vicariant_1978}
Rosen DE. 1978  Vicariant patterns and historical explanation in biogeography.
  {\em Systematic Zoology} \textbf{27}, 159.
(\href{http://dx.doi.org/10.2307/2412970}{10.2307/2412970})

\bibitem{Simberloff_there_1981}
Simberloff DS, Heck KL, McCoy ED, Connor EF. 1981  There Have Been No
  Statistical Tests of Cladistic Biogeographical Hypotheses!. In {\em
  Vicariance Biogeography: A Critique} ,  pp. 40--63. Columbia University
  Press.

\bibitem{Simberloff_calculating_1987}
Simberloff DS. 1987  Calculating Probabilities that Cladograms Match: {A}
  Method of Biogeographical Inference. {\em Systematic Zoology} \textbf{36},
  175.
(\href{http://dx.doi.org/10.2307/2413267}{10.2307/2413267})

\bibitem{savage_shape_1983}
Savage HM. 1983  The shape of evolution: systematic tree topology. {\em
  Biological Journal of the Linnean Society} \textbf{20}, 225--244.
(\href{http://dx.doi.org/10.1111/j.1095-8312.1983.tb01874.x}{10.1111/j.1095-8312.1983.tb01874.x})

\bibitem{slowinski_probabilities_1990}
Slowinski JB. 1990  Probabilities of $n$-trees under two models: {A}
  demonstration that asymmetrical interior nodes are not improbable. {\em
  Systematic Zoology} \textbf{39}, 89.
(\href{http://dx.doi.org/10.2307/2992212}{10.2307/2992212})

\bibitem{kirkpatrick_searching_1993}
Kirkpatrick M, Slatkin M. 1993  Searching for evolutionary patterns in the
  shape of a phylogenetic tree. {\em Evolution} \textbf{47}, 1171--1181.
(\href{http://dx.doi.org/10.1111/j.1558-5646.1993.tb02144.x}{10.1111/j.1558-5646.1993.tb02144.x})

\bibitem{mooers_tree_1995}
Mooers AO. 1995  Tree balance and tree completeness. {\em Evolution}
  \textbf{49}, 379--384.
(\href{http://dx.doi.org/10.1111/j.1558-5646.1995.tb02251.x}{10.1111/j.1558-5646.1995.tb02251.x})

\bibitem{mooers_inferring_1997}
Mooers AO, Heard SB. 1997  Inferring evolutionary process from phylogenetic
  tree shape. {\em The Quarterly Review of Biology} \textbf{72}, 31--54.
(\href{http://dx.doi.org/10.1086/419657}{10.1086/419657})

\bibitem{Aldous_stochastic_2001}
Aldous D. 2001  Stochastic models and descriptive statistics for phylogenetic
  trees, from {Y}ule to today. {\em Statistical Science} \textbf{16}.
(\href{http://dx.doi.org/10.1214/ss/998929474}{10.1214/ss/998929474})

\bibitem{blum_statistical_2005}
Blum MGB, Fran{\c{c}}ois O. 2005  On statistical tests of phylogenetic tree
  imbalance: the {S}ackin and other indices revisited. {\em Mathematical
  Biosciences} \textbf{195}, 141--153.
(\href{http://dx.doi.org/10.1016/j.mbs.2005.03.003}{10.1016/j.mbs.2005.03.003})

\bibitem{blum_which_2006}
Blum MGB, Fran{\c{c}}ois O. 2006  Which random processes describe the {Tree} of
  {Life}? {A} large-scale study of phylogenetic tree imbalance. {\em Systematic
  Biology} \textbf{55}, 685--691.

\bibitem{Verboom_species_2020}
Verboom GA, Boucher FC, Ackerly DD, Wootton LM, Freyman WA. 2020  Species
  selection regime and phylogenetic tree shape. {\em Systematic Biology}
  \textbf{69}, 774--794.
(\href{http://dx.doi.org/10.1093/sysbio/syz076}{10.1093/sysbio/syz076})

\bibitem{purvis_evaluating_2002}
Purvis A, Katzourakis A, Agapow PM. 2002  Evaluating phylogenetic tree shape:
  two modifications to {F}usco \& {C}ronk's method. {\em Journal of Theoretical
  Biology} \textbf{214}, 99--103.
(\href{http://dx.doi.org/10.1006/jtbi.2001.2443}{10.1006/jtbi.2001.2443})

\bibitem{fusco_new_1995}
Fusco G, Cronk QCB. 1995  A new method for evaluating the shape of large
  phylogenies. {\em Journal of Theoretical Biology} \textbf{175}, 235--243.
(\href{http://dx.doi.org/10.1006/jtbi.1995.0136}{10.1006/jtbi.1995.0136})

\bibitem{agapow_power_2002}
Agapow PM, Purvis A. 2002  Power of eight tree shape statistics to detect
  nonrandom diversification: a comparison by simulation of two models of
  cladogenesis. {\em Systematic Biology} \textbf{51}, 866--872.
(\href{http://dx.doi.org/10.1080/10635150290102564}{10.1080/10635150290102564})

\bibitem{heard_patterns_1992}
Heard SB. 1992  Patterns in tree balance among cladistic, phenetic, and
  randomly generated phylogenetic trees. {\em Evolution} \textbf{46},
  1818--1826.
(\href{http://dx.doi.org/10.1111/j.1558-5646.1992.tb01171.x}{10.1111/j.1558-5646.1992.tb01171.x})

\bibitem{heard_patternsrate_1996}
Heard SB. 1996  Patterns in phylogenetic tree balance with variable and
  evolving speciation rates. {\em Evolution} \textbf{50}, 2141--2148.

\bibitem{heard_shapes_2007}
Heard SB, Cox GH. 2007  The shapes of phylogenetic trees of clades, faunas, and
  local assemblages: exploring spatial pattern in differential diversification.
  {\em The American Naturalist} \textbf{169}, E107--E118.
(\href{http://dx.doi.org/10.1086/512690}{10.1086/512690})

\bibitem{kersting_genetic_2020}
Kersting SJ. 2020  Genetic programming as a means for generating improved tree
  balance indices ({M}aster's thesis, {U}niversity of {G}reifswald). .

\bibitem{kersting_measuring_2021}
Kersting SJ, Fischer M. 2021  Measuring tree balance using symmetry nodes - a
  new balance index and its extremal properties. {\em Mathematical Biosciences}
  \textbf{341}, 108690.
(\href{http://dx.doi.org/10.1016/j.mbs.2021.108690}{10.1016/j.mbs.2021.108690})

\bibitem{sackin_good_1972}
Sackin MJ. 1972  ``{G}ood'' and ``bad'' phenograms. {\em Systematic Biology}
  \textbf{21}, 225--226.
(\href{http://dx.doi.org/10.1093/sysbio/21.2.225}{10.1093/sysbio/21.2.225})

\bibitem{colless_review_1982}
Colless D. 1982  Review of ``{P}hylogenetics: the theory and practice of
  phylogenetic systematics''. {\em Systematic Zoology} \textbf{31}, 100--104.

\bibitem{coronado_balance_2019}
Coronado TM, Mir A, Rossell{\'{o}} F, Valiente G. 2019  A balance index for
  phylogenetic trees based on rooted quartets. {\em Journal of Mathematical
  Biology} \textbf{79}, 1105--1148.
(\href{http://dx.doi.org/10.1007/s00285-019-01377-w}{10.1007/s00285-019-01377-w})

\bibitem{FischerKerstingWickeSupp}
Kersting SJ, Wicke K, Fischer M. 2024  Tree balance in phylogenetic models:
  {S}upplementary material.. \url{https://rb.gy/8a3kfg},
  \url{https://github.com/MareikeFischer/powerRbalSUPP/blob/main/SUPPLEMENT_Tree_balance_in_phylogenetic_models.pdf}.

\bibitem{mooers_models_2007}
Mooers A, Harmon L, Blum M, Wong D, Heard S. 2007 pp.~--.
In {\em Some models of phylogenetic tree shape}, pp.~--. Oxford University
  Press.

\bibitem{khurana_limits_2023}
Khurana MP, Scheidwasser-Clow N, Penn MJ, Bhatt S, Duch\^ene DA. 2023  The
  Limits of the Constant-rate Birth–Death Prior for Phylogenetic Tree
  Topology Inference. {\em Systematic Biology}.
(\href{http://dx.doi.org/10.1093/sysbio/syad075}{10.1093/sysbio/syad075})

\bibitem{matsen_geometric_2006}
Matsen FA. 2006  A geometric approach to tree shape statistics. {\em Systematic
  Biology} \textbf{55}, 652--661.
(\href{http://dx.doi.org/10.1080/10635150600889617}{10.1080/10635150600889617})

\bibitem{hayati_new_2019}
Hayati M, Shadgar B, Chindelevitch L. 2019  A new resolution function to
  evaluate tree shape statistics. {\em {PLOS} {ONE}} \textbf{14}, e0224197.
(\href{http://dx.doi.org/10.1371/journal.pone.0224197}{10.1371/journal.pone.0224197})

\bibitem{williamson_genealogy_2002}
Williamson S, Orive ME. 2002  {The genealogy of a sequence subject to purifying
  selection at multiple sites}. {\em Molecular {B}iology and {E}volution}
  \textbf{19}, 1376--1384.
(\href{http://dx.doi.org/10.1093/oxfordjournals.molbev.a004199}{10.1093/oxfordjournals.molbev.a004199})

\bibitem{shao_tree_1990}
Shao KT, Sokal RR. 1990  Tree balance. {\em Systematic Zoology} \textbf{39},
  266.
(\href{http://dx.doi.org/10.2307/2992186}{10.2307/2992186})

\bibitem{furnas_generation_1984}
Furnas GW. 1984  The generation of random, binary unordered trees. {\em Journal
  of Classification} \textbf{1}, 187--233.
(\href{http://dx.doi.org/10.1007/bf01890123}{10.1007/bf01890123})

\bibitem{colijn_metric_2018}
Colijn C, Plazzotta G. 2018  A metric on phylogenetic tree shapes. {\em
  Systematic Biology} \textbf{67}, 113--126.
(\href{http://dx.doi.org/10.1093/sysbio/syx046}{10.1093/sysbio/syx046})

\bibitem{rosenberg_colijn_2020}
Rosenberg NA. 2021  On the {C}olijn–{P}lazzotta numbering scheme for
  unlabeled binary rooted trees. {\em Discrete Applied Mathematics}
  \textbf{291}, 88--98.
(\href{http://dx.doi.org/10.1016/j.dam.2020.11.021}{10.1016/j.dam.2020.11.021})

\bibitem{herrada_scaling_2011}
Herrada EA, Egu{\'{\i}}luz VM, Hern{\'{a}}ndez-Garc{\'{\i}}a E, Duarte CM. 2011
   Scaling properties of protein family phylogenies. {\em {BMC} Evolutionary
  Biology} \textbf{11}.
(\href{http://dx.doi.org/10.1186/1471-2148-11-155}{10.1186/1471-2148-11-155})

\bibitem{ford_probabilities_2005}
Ford DJ. 2005  Probabilities on cladograms: introduction to the alpha model. .

\bibitem{hernandez_scaling_2010}
Hern\'{a}ndez-Garc\'{i}a E, Tu\u{g}rul M, Alejandro~Herrada E, Egu\'{i}luz VM,
  Klemm K. 2010  Simple models for scaling phylogenetic trees. {\em
  International Journal of Bifurcation and Chaos} \textbf{20}, 805–811.
(\href{http://dx.doi.org/10.1142/s0218127410026095}{10.1142/s0218127410026095})

\bibitem{mir_sound_2018}
Mir A, Rotger L, Rossell{\'{o}} F. 2018  Sound {C}olless-like balance indices
  for multifurcating trees. {\em {PLOS} {ONE}} \textbf{13}, e0203401.
(\href{http://dx.doi.org/10.1371/journal.pone.0203401}{10.1371/journal.pone.0203401})

\bibitem{bartoszek_squaring_2021}
Bartoszek K, Coronado TM, Mir A, Rossell{\'{o}} F. 2021  Squaring within the
  {C}olless index yields a better balance index. {\em Mathematical Biosciences}
  \textbf{331}, 108503.
(\href{http://dx.doi.org/10.1016/j.mbs.2020.108503}{10.1016/j.mbs.2020.108503})

\bibitem{rogers_central_1996}
Rogers JS. 1996  Central moments and probability distributions of three
  measures of phylogenetic tree imbalance. {\em Systematic Biology}
  \textbf{45}, 99--110.
(\href{http://dx.doi.org/10.1093/sysbio/45.1.99}{10.1093/sysbio/45.1.99})

\bibitem{mir_new_2013}
Mir A, Rossell{\'{o}} F, Rotger L. 2013  A new balance index for phylogenetic
  trees. {\em Mathematical Biosciences} \textbf{241}, 125--136.
(\href{http://dx.doi.org/10.1016/j.mbs.2012.10.005}{10.1016/j.mbs.2012.10.005})

\bibitem{knuth_volume3_1998}
Knuth DE. 1998 {\em The art of computer programming volume 3: Sorting and
  searching}.
Addison-Wesley Professional 2nd edition.

\bibitem{dobrow_fill_1999}
Dobrow RP, Fill JA. 1999  Total Path Length for Random Recursive Trees. {\em
  Combinatorics, Probability and Computing} \textbf{8}, 317–333.
(\href{http://dx.doi.org/10.1017/S0963548399003855}{10.1017/S0963548399003855})

\bibitem{takacs_1992}
Takacs L. 1992  On the total heights of random rooted trees. {\em Journal of
  Applied Probability} \textbf{29}, 543–556.
(\href{http://dx.doi.org/10.2307/3214892}{10.2307/3214892})

\bibitem{takacs_1994}
Takacs L. 1994  On the Total Heights of Random Rooted Binary Trees. {\em
  Journal of Combinatorial Theory, Series B} \textbf{61}, 155--166.
(\href{http://dx.doi.org/10.1006/jctb.1994.1041}{10.1006/jctb.1994.1041})

\bibitem{coronado_sackins_2020}
Coronado TM, Mir A, Rossell{\'{o}} F, Rotger L. 2020  On {S}ackin's original
  proposal: the variance of the leaves' depths as a phylogenetic balance index.
  {\em {BMC} Bioinformatics} \textbf{21}.
(\href{http://dx.doi.org/10.1186/s12859-020-3405-1}{10.1186/s12859-020-3405-1})

\bibitem{mckenzie_distributions_2000}
McKenzie A, Steel M. 2000  Distributions of cherries for two models of trees.
  {\em Mathematical Biosciences} \textbf{164}, 81--92.
(\href{http://dx.doi.org/10.1016/s0025-5564(99)00060-7}{10.1016/s0025-5564(99)00060-7})

\bibitem{yule_mathematical_1925}
Yule GU. 1925  A mathematical theory of evolution, based on the conclusions of
  {D}r. {J}. {C}. {W}illis, {F}. {R}. {S}. {\em Philosophical Transactions of
  the Royal Society of London. Series B, Containing Papers of a Biological
  Character} \textbf{213}, 21--87.
(\href{http://dx.doi.org/10.1098/rstb.1925.0002}{10.1098/rstb.1925.0002})

\bibitem{Steel2001properties}
Steel M, McKenzie A. 2001  Properties of phylogenetic trees generated by
  {Y}ule-type speciation models. {\em Mathematical Biosciences} \textbf{170},
  91--112.

\bibitem{steel_phylogeny_2016}
Steel M. 2016 {\em Phylogeny: Discrete and random processes in evolution}.
{CBMS}-{NSF} regional conference series in applied mathematics. Philadelphia
  PA: Society for Industrial and Applied Mathematics.

\bibitem{Fuchs2024}
Fuchs M. 2024  Shape Parameters of Evolutionary Trees in Theoretical Computer
  Science. {\em Philosophical Transactions of the Royal Society B (submitted)}.

\bibitem{Dickey2024}
Dickey EH, Rosenberg NA. 2024  Labeled histories with multifurcation and
  simultaneity. {\em Philosophical Transactions of the Royal Society B
  (submitted)}.

\bibitem{king_mathematical_2023}
King MC, Rosenberg NA. 2023  A Mathematical Connection Between
  Single-Elimination Sports Tournaments and Evolutionary Trees. {\em
  Mathematics Magazine} \textbf{96}, 484–497.
(\href{http://dx.doi.org/10.1080/0025570x.2023.2266389}{10.1080/0025570x.2023.2266389})

\bibitem{aldous_random_1996}
Aldous D. 1996  Probability {D}istributions on {C}ladograms. In {\em Random
  Discrete Structures} ,  pp. 1--18. Springer New York.
(\href{http://dx.doi.org/10.1007/978-1-4612-0719-1\_1}{10.1007/978-1-4612-0719-1\_1})

\bibitem{harvey_phylogenies-without-fossils_1994}
Harvey PH, May RM, Nee S. 1994  Phylogenies Without Fossils. {\em Evolution}
  \textbf{48}, 523--529.
(\href{http://dx.doi.org/10.2307/2410466}{10.2307/2410466})

\bibitem{maddison_BiSSE_2007}
Maddison WP, Midford PE, Otto SP. 2007  Estimating a binary character’s
  effect on speciation and extinction. {\em Systematic Biology} \textbf{56},
  701–710.
(\href{http://dx.doi.org/10.1080/10635150701607033}{10.1080/10635150701607033})

\bibitem{fitzjohn_QuaSSE_2010}
FitzJohn RG. 2010  Quantitative Traits and Diversification. {\em Systematic
  Biology} \textbf{59}, 619–633.
(\href{http://dx.doi.org/10.1093/sysbio/syq053}{10.1093/sysbio/syq053})

\bibitem{fitzjohn_MuSSE_2012}
FitzJohn RG. 2012  Diversitree: comparative phylogenetic analyses of
  diversification in {R}. {\em Methods in Ecology and Evolution} \textbf{3},
  1084–1092.
(\href{http://dx.doi.org/10.1111/j.2041-210x.2012.00234.x}{10.1111/j.2041-210x.2012.00234.x})

\bibitem{goldberg_GeoSSE_2011}
Goldberg EE, Lancaster LT, Ree RH. 2011  Phylogenetic Inference of Reciprocal
  Effects between Geographic Range Evolution and Diversification. {\em
  Systematic Biology} \textbf{60}, 451–465.
(\href{http://dx.doi.org/10.1093/sysbio/syr046}{10.1093/sysbio/syr046})

\bibitem{freyman_ChromoSSE_2017}
Freyman WA, H\"{o}hna S. 2017  Cladogenetic and Anagenetic Models of Chromosome
  Number Evolution: {A} Bayesian Model Averaging Approach. {\em Systematic
  Biology} \textbf{67}, 195–215.
(\href{http://dx.doi.org/10.1093/sysbio/syx065}{10.1093/sysbio/syx065})

\bibitem{vasconcelos_HiSSE_2022}
Vasconcelos T, O’Meara BC, Beaulieu JM. 2022  A flexible method for
  estimating tip diversification rates across a range of speciation and
  extinction scenarios. {\em Evolution} \textbf{76}, 1420–1433.
(\href{http://dx.doi.org/10.1111/evo.14517}{10.1111/evo.14517})

\bibitem{kaur_distributions_2023}
Kaur G, Choi KP, Wu T. 2023  Distributions of cherries and pitchforks for the
  {F}ord model. {\em Theoretical Population Biology} \textbf{149}, 27--38.

\bibitem{paradis_enumeration_2023}
Paradis E. 2023 {\em Enumeration and simulation of random tree topologies}.

\bibitem{innan_statistical_2005}
Innan H, Zhang K, Marjoram P, Tavar{\'e} S, Rosenberg NA. 2005  Statistical
  tests of the coalescent model based on the haplotype frequency distribution
  and the number of segregating sites. {\em Genetics} \textbf{169}, 1763--1777.

\bibitem{semple_phylogenetics_2003}
Semple C, Steel M. 2003 {\em Phylogenetics ({O}xford lecture series in
  mathematics and its applications)}.
Oxford University Press.

\bibitem{bienvenu_revisiting_2020}
Bienvenu F, Cardona G, Scornavacca C. 2021  Revisiting {S}hao and {S}okal’s
  {B}2 index of phylogenetic balance. {\em Journal of Mathematical Biology}
  \textbf{83}.
(\href{http://dx.doi.org/10.1007/s00285-021-01662-7}{10.1007/s00285-021-01662-7})

\bibitem{fischer_tree_2021}
Fischer M, Herbst L, Kersting SJ, K{\"u}hn L, Wicke K. 2021  Tree balance
  indices: a comprehensive survey. .

\bibitem{marquitti_allopatry_2020}
Marquitti FMD, Fernandes LD, de~Aguiar MAM. 2020  Allopatry increases the
  balance of phylogenetic trees during radiation under neutral speciation. {\em
  Ecography} \textbf{43}, 1487--1498.
(\href{http://dx.doi.org/10.1111/ecog.04937}{10.1111/ecog.04937})

\bibitem{rohlf_accuracy_1990}
Rohlf FJ, Chang WS, Sokal RR, Kim J. 1990  Accuracy of estimated phylogenies:
  effects of tree topology and evolutionary model. {\em Evolution} \textbf{44}.

\bibitem{huelsenbeck_phylogenetic_1996}
Huelsenbeck JP, Kirkpatrick M. 1996  Do phylogenetic methods produce trees with
  biased shapes?. {\em Evolution} \textbf{50}, 1418--1424.
(\href{http://dx.doi.org/10.1111/j.1558-5646.1996.tb03915.x}{10.1111/j.1558-5646.1996.tb03915.x})

\bibitem{blum_matrilineal_2006}
Blum MGB, Heyer E, Fran{\c{c}}ois O, Austerlitz F. 2006  Matrilineal fertility
  inheritance detected in hunter-gatherer populations using the imbalance of
  gene genealogies. {\em {PLoS} Genetics} \textbf{2}, e122.
(\href{http://dx.doi.org/10.1371/journal.pgen.0020122}{10.1371/journal.pgen.0020122})

\bibitem{Aldous_1991_ContI}
Aldous D. 1991a  The continuum random tree {I}. {\em The Annals of Probability}
  \textbf{19}, 1--28.
(\href{http://dx.doi.org/10.1214/aop/1176990534}{10.1214/aop/1176990534})

\bibitem{Aldous_1991_ContII}
Aldous D. 1991b  The continuum random tree {II}: {A}n overview. In {\em
  Stochastic Analysis} ,  p. 23–70. Cambridge University Press.
(\href{http://dx.doi.org/10.1017/cbo9780511662980.003}{10.1017/cbo9780511662980.003})

\bibitem{Aldous_1993_ContIII}
Aldous D. 1993  The continuum random tree {III}. {\em The Annals of
  Probability} \textbf{21}, 248--289.
(\href{http://dx.doi.org/10.1214/aop/1176989404}{10.1214/aop/1176989404})

\end{thebibliography}

\end{document}